\documentclass{llncs}
\usepackage{amsmath, amssymb}
\usepackage{graphicx}
\usepackage{subfig}
\usepackage{wrapfig}
\usepackage{tabularx}
\usepackage{tikz}
\usepackage{multirow}
\usepackage{ascii}
\usepackage{paralist}
\usepackage{enumerate}
\usepackage{times}

\newif\ifreview
\reviewfalse

\ifreview
\usepackage{setspace}
\usepackage{lineno}
\linenumbers
\fi

\newcommand*\circled[1]{
  \protect\tikz[baseline=(char.base)]{
    \protect\node[shape=circle,draw,inner sep=0.2pt] (char) {#1};}}

\newcommand{\yes}{\ensuremath{\circled{\checkmark}}}
\newcommand{\no}{\ensuremath{\circled{$\times$}}}

\newcommand{\algo}[1]{{\normalfont\scshape #1}}
\DeclareMathAlphabet{\mathcal}{OMS}{cmsy}{m}{n}

\title{Clustering Evolving Networks}
\author{
	Tanja Hartmann\inst{},
	Andrea Kappes,
	Dorothea Wagner
}
\institute{
  Department of Informatics, Karlsruhe Institute of Technology (KIT)
\\ \url{\{tanja.hartmann, andrea.kappes, dorothea.wagner\}@kit.edu}
}
\begin{document}
\setcounter{totalnumber}{8}
\setcounter{topnumber}{8}
\setcounter{bottomnumber}{8}
\renewcommand{\textfraction}{0.001}
\renewcommand{\topfraction}{0.999}
\renewcommand{\bottomfraction}{0.999}
\maketitle
%
\vspace{-0ex}
\begin{abstract}
  Roughly speaking, clustering evolving networks aims at detecting
  structurally dense subgroups in networks that evolve over time.
  This implies that the subgroups we seek for also evolve, which
  results in many additional tasks compared to clustering static
  networks.  We discuss these additional tasks and difficulties
  resulting thereof and present an overview on current approaches to
  solve these problems.  We focus on clustering approaches in online
  scenarios, i.e., approaches that incrementally use structural
  information from previous time steps in order to incorporate
  temporal smoothness or to achieve low running time.  Moreover, we
  describe a collection of real world networks and generators for
  synthetic data that are often used for evaluation.
\end{abstract}

\section{Introduction}
\label{sec:introduction}
Clustering is a powerful tool to examine the structure of various
data.  Since in many fields data often entails an inherent network
structure or directly derives from physical or virtual networks,
clustering techniques that explicitly build on the information given
by links between entities recently received great attention. Moreover,
many real world networks are continuously evolving, which makes it
even more challenging to explore their structure.  Examples for
evolving networks include networks based on mobile communication data,
scientific publication data, and data on human interaction.

The structure that is induced by the entities of a network together
with the links between is often called \emph{graph}, the entities are
called \emph{vertices} and the links are called \emph{edges}.
However, the terms graph and network are often used
interchangeably. The structural feature that is classically addressed
by graph clustering algorithms are subsets of vertices that are linked
significantly stronger to each other than to vertices outside the
subset.  In the context of mobile communication networks this could
be, for example, groups of cellphone users that call each other more
frequently than others.
Depending on the application and the type of the underlying network,
searching for this kind of subsets has many different names.
Sociologists usually speak about \emph{community detection} or
\emph{community mining} in social networks, in the context of
communication networks like Twitter, people aim at \emph{detecting
  emerging topics} while in citations networks the focus is on the
\emph{identification of research areas}, to name but a few.  All these
issues can be solved by modeling the data as an appropriate graph and
applying graph clustering.
The found sets (corresponding to communities, topics or research
areas) are then called \emph{clusters} and the set of clusters is
called a \emph{clustering}.  
We further remark that also beyond sociology the term~\emph{community}
is often used instead of cluster~\cite{n-dcsn-04}.
The notion of clusters or communities as densely
connected subgroups that are only sparsely connected to each other has
led to the paradigm of intracluster density versus intercluster
sparsity in the field of graph clustering.  

Nevertheless, the notion of a clustering given so far still leaves
room for many different formal definitions.
Most commonly, a clustering of a graph is defined as a partition of
the vertex set into subsets, which form the clusters.
In some scenarios (e.g., outlier detection) it is, however,
undesirable that each vertex is assigned to a cluster.  In this case,
a clustering not necessarily forms a partition of the vertex set, but
leaves some vertices \emph{unclustered}.  Yet both concepts are based
on disjoint vertex sets, and the latter can be easily transformed into
the former one by just considering each vertex that is not a member of
a cluster as a cluster consisting of exactly one vertex.  Other applications further admit
overlapping clusters, again with or without a complete assignment of
the vertices to clusters.

\begin{figure}[tb]
  \centering
  \includegraphics[page=2]{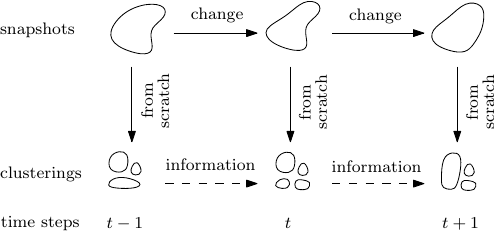}
  \caption{Evolutionary clustering strategy in evolving graphs.
    Horizontal dashed arrows indicate the use of information, vertical
    arrows indicate the evolutionary strategy based on a static
    clustering approach applied from scratch.}
  \label{fig:evolClus}
\end{figure}

\begin{figure}[tb]
  \centering
  \includegraphics[page =1]{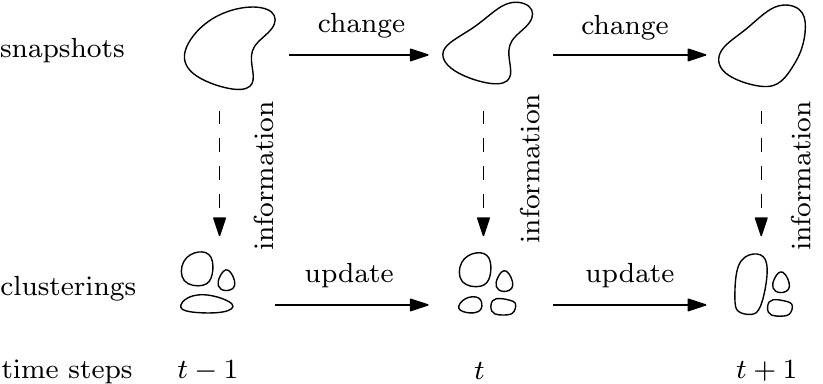}
  \caption{Dynamic update strategy for clusterings in evolving graphs.
    Vertical dashed arrows indicate the use of information, horizontal
    arrows indicate the dynamic update strategy based on previous time
    steps.}
  \label{fig:dynClus}
\end{figure}

In this survey we give an overview of recent graph clustering
approaches that aim at finding disjoint or overlapping clusters in
evolving graphs.  The evolution of the graph is usually modeled
following one of two common concepts: The first concept is based on a
series of snapshots of the graph, where each snapshot corresponds to a
time step, and the difference between two consecutive snapshots
results from a bunch of edge and vertex changes.  The second concept
considers a given stream of atomic edge and vertex changes, where each
change induces a new snapshot and a new time step.  The primary
objective of clustering such networks is to find a meaningful
clustering for each snapshot.  Some algorithms further aim at a
particularly fast computation of these clusterings, others assume that
changes have only a small impact on the community structure in each
time step, and thus, aim at clusterings that differ not too much in
consecutive time steps.  The latter was introduced as \emph{temporal
  smoothness} by Chakrabarti et al.~\cite{ckt-ec-06} in the context of
clustering evolving attributed data (instead of graphs).
In order to achieve these goals, \emph{online} algorithms explicitly
exploit information about the graph structure and the community
structure of previous time steps.  Algorithms that further use
structural information from following time steps are called
\emph{offline}.  In this survey, we consider only online algorithms
that can be roughly separated into two classes.
The first class contains clustering approaches that incorporate
temporal smoothness inspired by Chakrabarti et al.  Most of these
approaches are based on an existing static clustering algorithm, which
is executed from scratch in each time step
(cp.~Figure~\ref{fig:evolClus}).  In contrast, the approaches in the
second class dynamically update clusterings found in previous time
steps without a computation from scratch
(cp.~Figure~\ref{fig:dynClus}).
%

Apart from finding an appropriate clustering in each snapshot of an
evolving graph, many applications require further steps in order to
make the found clusterings interpretable and usable for further
analysis.  A first natural question directly resulting from the
evolution of the graph is how the found clusters or communities evolve
over time and at what time steps events like cluster merging or
cluster splitting occur.  In order to answer this question, the
clusters need to be tracked over time, thereby
finding sequences of snapshots where certain clusters remain stable
while other clusters may split or merge.
In this context, clusters or communities of a single snapshot are
often called \emph{local} in order to distinguish them from sequences
of associated (local) communities in consecutive snapshots, which
describe the evolution of a certain
\emph{(meta)community} over time.
When the evolution of the clusters is supposed to be interpreted by
human experts, 
it is further necessary to present the algorithmic results in a clear
and readable form. Hence, the visualization of evolving clusters is
another central issue in the context of clustering evolving graphs.
The evaluation of found clusterings is finally an issue
regarding the design of good clustering algorithms.  There are
many open questions on how to choose an appropriate evaluation scheme
in order to get credible and comparable results.  We discuss these
issues in more detail in Section~\ref{subsec:mainIssues} 
and give a brief idea on applications based on clustering evolving
graphs in Section~\ref{subsec:app}.

\paragraph{Delimitation.}
Apart from clustering approaches that follow the intracluster density
versus intercluster sparsity paradigm, there exist various further
approaches that look very similar at a first glance but turn out to
have a different focus.

Very closely related to graph clustering are algorithms for
\emph{graph partitioning}~\cite{bs-gp-11}.  In contrast to many graph
clustering algorithms, graph partitioning always assumes that the
number of clusters is an input parameter, most often, a power of $2$,
and seeks to minimize the number of edges cut by a partition, such
that the parts have (almost) equal size.  Its main application is not
network analysis but the preprocessing of graphs for parallel
computing tasks.  The dynamic counterpart to static graph
partitioning is often called \emph{repartitioning} or \emph{load
  balancing}~\cite{cbdbhr-la-07,c-d-89,m-dlbpn-09}.
Another area similar to clustering evolving graphs is clustering
\emph{graph streams}~\cite{azy-a-10,zy-ogscs-13}.  Similar to
consecutive graph snapshots, a graph stream is a sequence of
consecutively arriving graphs, but instead of finding a clustering of
the vertices in each graph or snapshot, the aim is to detect groups of
similar graphs.  The term \emph{streaming algorithm} is usually used
for algorithms that process the data in one or few passes under the
restriction of limited memory availability, like for example the
partitioning algorithm by Stanton and Kliot~\cite{sk-sgpld-12}.
However, some authors also use the adjective~\emph{streaming} or the
term~\emph{stream model} in the context of graph changes in order to
describe consecutive atomic changes~\cite{asks-d-12,dyll-ik-12}.  A
further task is the search for stable subgraphs in a given time
interval in an evolving network, i.e., subgraphs that change only
slightly during the whole interval.  Depending on the formal
definition of stability these subgraphs have various other names, like
\emph{heavy subgraphs} or \emph{high-score
  subgraphs}~\cite{bms-mhste-11}.  \emph{Pattern Mining} in evolving
graphs is focused on frequently occurring subgraphs, independent from
their density~\cite{bkw-pmfds-06}. 


\paragraph{Intention and Outline.}
In this survey, we introduce some of the current graph clustering
approaches for evolving networks that operate in an online scenario.
All approaches have in common that they use structural information
from the previous time steps in order to generate a meaningful
clustering for the snapshot of the current time step.  In doing so,
some approaches focus on temporal smoothness, while other approaches
aim at a fast running time and a few even achieve both.

In contrast to existing surveys on graph clustering, we focus on
online algorithms in evolving networks.  A very detailed and
well-founded presentation of algorithmic aspects in static
(non-evolving) and evolving graphs is further given in the theses of
G\"orke~\cite{g-aawsd-10}.  For an overview on
clustering techniques in static graphs see also
Schaeffer~\cite{s-gc-07} and Fortunato et al.~\cite{f-c-09}.  The
latter also provide a short abstract on clustering evolving graphs.
Aynaud et al.~\cite{afgw-cendd-13} explicitly consider clustering
approaches in evolving graphs, however, they do not focus on the
algorithmic aspect of reusing structural information in an online
scenario.  Finally, Bilgin and Yener~\cite{by-dnemc-08} consider
evolving networks from a more general perspective. They also provide a
section on "Clustering Dynamic Graphs'', however, the emphasis of this
section is the above mentioned idea of clustering graph streams.

This paper is organized as follows.  In
Section~\ref{subsec:mainIssues}, we discuss the above mentioned main
issues related to clustering evolving networks in more detail.  In
Section~\ref{subsec:measures} we provide an overview on popular
quality and distance measures for clusterings.  The former
are of course used for the evaluation of clusterings but also in the
context of algorithm design.  The latter are used for the evaluation
as well as for cluster tracking and event detection.  We conclude the
introduction by a a brief idea on applications in
Section~\ref{subsec:app}.  The main part of
this survey is presented in Section~\ref{sec:IncrClus}, where we
introduce current clustering approaches according to our focus
described above.  Moreover, we provide an overview on the main
features of the presented algorithms in
Table~\ref{tab:properties_dyn}.  Section~\ref{sec:dataSets} further
lists a selection of data sets and graph generators used for the
evaluation of the approaches presented in Section~\ref{sec:IncrClus}
and briefly discusses the difficulties in choosing appropriate data
for evaluating clustering approaches on evolving graphs.  We finally
conclude in Section~\ref{sec:conclusion}.

\paragraph{Notation.}
Until noted otherwise, we will assume that graphs are \emph{simple}, i.e., they do not contain loops and parallel edges.
A \emph{dynamic} or \emph{evolving} graph $\mathcal{G} = (G_0, \ldots, G_{t_\text{max}})$ is a sequence of graphs with $G_t = (V_t, E_t)$ being the state of the dynamic graph at \emph{time step} $t$.
$G_t$ is also called \emph{snapshot} of $\mathcal{G}$.
A \emph{clustering} $\mathcal{C}_t = \{C_1, \ldots, C_k\}$ of $G_t$ is a set of subsets of $V_t$ called \emph{clusters} or \emph{communities}.
If these subsets are pairwise disjoint, the clustering is called \emph{disjoint}, otherwise it is \emph{overlapping}.
A disjoint clustering that further has the property that each vertex is contained in a cluster, i.e., that corresponds to a partition of $V_t$, is called \emph{complete}.
Complete clusterings are often represented by storing a \emph{cluster id} for each vertex that encodes the corresponding cluster.
A pair $\{u,v\}$ of vertices such that there is a cluster that contains both $u$ and $v$ is called \emph{intracluster pair}, otherwise $\{u,v\}$ is called \emph{intercluster pair}.
An edge between the vertices of an intracluster pair is called \emph{intracluster edge}; \emph{intercluster edges} are defined analogously.
A \emph{singleton clustering} is a complete clustering where each cluster contains only one vertex; such clusters are called \emph{singleton clusters}.
The other extreme, i.e., a clustering consisting of only one cluster
containing all vertices, is called \emph{1-clustering}.
Each cluster~$C\subset V_t$ further induces a \emph{cut} in~$G_t$.
A cut in a graph~$G= (V,E)$ is defined by a set~$S\subset V$,
which indicates one side of the cut.  The other side is implicitly
given by~$V\setminus S$.  A cut is thus denoted by~$(S,V\setminus
S)$.

\subsection{Main Issues When Clustering Evolving Networks
}\label{subsec:mainIssues}
In the following we briefly discuss the main issues related to
clustering evolving networks.  We consider cluster tracking and
visualization first, since these problems can be solved independent
from the cluster detection.  Our remarks on cluster detection in
online scenarios give a rough idea on different techniques used in
this field, followed by a short overview on some state-of-the-art
evaluation techniques.

\paragraph{Cluster Tracking and Event Detection.}
Assuming the cluster structure of the network is already given for
each snapshot by an arbitrary clustering approach, detecting the
evolution of the clusters over time becomes a task independent from
finding the clusters.  Most approaches that address this task describe
a framework of two subproblems.  On the one hand, they seek for series
of similar clusters in consecutive snapshots (often called meta
communities, meta groups or time-lines), and on the other hand, they
aim at identifying critical events where clusters, for instance,
appear, survive, disappear, split or merge.  In particular, deciding
if a cluster that has just disappeared reappears in future time steps, and
thus, actually survives,
requires future information, which is not available in an online
scenario.  Hence, whether a framework is applicable in an online
scenario depends on the defined events.  The frameworks of Takaffoli
et al.~\cite{tfsz-t-11} and Green et al.~\cite{gdc-tecds-10} are
offline frameworks since they compare the structure of the clusters of
the current snapshot to all previous and future snapshots in order to
also find clusters that disappear and reappear after a while.  While
the framework by Takaffoli et al.\ requires disjoint clusters, the
approach of Green et al.\ also allows overlapping clusters.

In order to compare clusters of consecutive snapshots, many
approaches define a similarity measure considering two clusters as
similar if the similarity value exceeds a given threshold.  Asur et
al.~\cite{apu-a-09} detect similar clusters by the size of their
intersection.  Takaffoli et al.~\cite{tfsz-t-11} use the size of the
intersection over the size of the larger cluster.  Berger-Wolf and
Saia~\cite{bs-afads-06} admit the use of any similarity measure that
is efficiently computable and satisfies some mathematical properties,
as it does the standard Jaccard similarity measure~\cite{j-t-12} that
describes the size of the intersection over the size of the union.
 
The frameworks mentioned so far can be applied to any cluster
structure in a given graph, regardless which clustering approach was
used to find this structure.  Other tracking approaches, however,
exploit special properties of the given cluster structures in the
snapshots, and thus, require that the clusters are constructed by a
designated (static) clustering method.  Palla et al.~\cite{pbv-q-07}
require clusterings found by the clique percolation method
(PCM)~\cite{dpv-cprn-05}, which can be also seen as a special case of a
clustering method proposed by Everett and Borgatti~\cite{eb-aco-98}.
For a brief description of PCM see the part in
Section~\ref{sec:IncrClus} where algorithms maintaining auxiliary
structures are introduced.  In order to identify evolving clusters
in two consecutive time steps, Palla et al.\ construct the union of
the two corresponding snapshots and apply again PCM to this union
graph.  Due to the properties of the clusters found by PCM, it
holds that each cluster found in one of the snapshots is contained
in exactly one cluster in the union graph.  A cluster~$C$ in
the snapshot at time~$t-1$ is then associated with the cluster~$C'$
in the snapshot at time~$t$ that is contained in the same cluster in
the union graph and has the most vertices in common with~$C$.
Falkowski et al.~\cite{fbs-mvess-06} consider clusters that result
from a hierarchical divisive edge betweenness clustering algorithm.
In contrast to Palla et al.\ who map clusters only between two
consecutive time steps, Falkowski et al.\ present an offline approach.
They construct an auxiliary graph that consists of all clusters found
in any snapshot and two clusters are connected by an edge if and only
if the relative overlap of both clusters exceeds a given threshold.
On this graph the authors apply the same clustering algorithm as on
the snapshots in order to find groups of (local) communities that are
similar in different time steps.  Obviously, this approach is an
offline approach.
Another offline approach is given by Tantipathananandh et
al.~\cite{bkt-affci-07}.  Here the authors assume given groups in the
snapshots, where members of the same group interact while members of
different groups do not interact.  Based on observed interactions of
entities in consecutive time steps, an auxiliary graph is built on
which the authors solve a coloring problem.  Finally, there are also
some frameworks that are not especially designed for networks, but are
general enough to be applied to networks as well~\cite{snts-monic-06}.
For a further categorization of tracking methods see also Aynaud et
al.~\cite{afgw-cendd-13}.

Besides whole frameworks that track clusters according to similarity,
tracking clusters by cluster ids is a very natural and simple
approach.  This however requires that clusters that are given the same
id in two consecutive time steps are somehow related.  Many graph
clustering algorithms that dynamically update previous clusterings,
like \algo{Label Propagation}~\cite{pcw-a-09},
\algo{LabelRankT}~\cite{xcs-lrtic-13} and
\algo{DiDic}~\cite{gm-addhc-10}, to name but a few of the approaches
introduced in Section~\ref{sec:IncrClus}, fulfill this requirement.
The relation between clusters of the same id depends on how the
particular algorithm chooses the cluster ids.  Furthermore, inferring
clusters based on generative models as done by
\algo{FacetNet}~\cite{lczst-a-09} often admits the tracking of
clusters without an additional framework, as the evolution of the
clusters can be read off the resulting model.  

\paragraph{Visualization of the Evolution of Clusters.}
Even if there is already a clustering given for each snapshot and we
know which local clusters correspond to each other in different
snapshots, the visualization of the evolution of these clusters is
still not trivial.  Since the evolution of clusters involves not only
one graph but several snapshots of an evolving graph, the total object
that needs to be visualized quickly becomes very complex.  Hence, one
problem that needs to be dealt with when visualizing such an object is
simplification.  Many visualization approaches and in particular
interactive visualization tools solve this problem by offering
different views on different aspects of the
object~\cite{cfgrstvz-mcmds-10,fbs-mvess-06,smm-cvnld-13}.  Apart from
different views, the most intuitive layout for evolving clusterings is
probably to draw consecutive snapshots next to each other and depict
the correspondence of clusters for example by colors or by additional
edges between the snapshots.  In this kind of layout another popular
task in the field of visualization gains importance, namely the
preservation of the mental map.  This means that corresponding
clusters (and also vertices) in different snapshots are placed at
similar positions in the image of the particular snapshot, such that
potential changes in the clustering structure can be easily
recognized.  The goal to generate well readable diagrams further also
justifies the postulation of temporal smoothness.  Instead of drawing
the snapshots next to each other, TeCFlow~\cite{gz-tcfat-04} and
SoNIA~\cite{mmb-dnv-05} provide the possibility to create little
movies out of consecutive snapshots.  Moreover, many visualization
approaches are proposed on top of new clustering approaches or
existing tracking frameworks~\cite{hk-av-13}.  However, these
approaches are often specially tailored with respect to these
clustering algorithms or tracking frameworks.

\paragraph{Online Cluster Detection in Evolving Graphs.}
The most intuitive attempt to deal with clusters in evolving graphs is
to cluster each snapshot independently with a static clustering
algorithm and track the clusters afterwards in order to uncover their
evolution.  However, depending on the static clustering algorithm that
is used to find the clusters, the structure of the clusters in each
snapshot may vary greatly such that a tracking method may find a lot
of change points instead of nicely evolving clusters.  Greedy
agglomerative approaches, for example, that aim at optimizing an
objective function often tend to find different local optima in
consecutive snapshots, depending on the order in which the vertices
are chosen.
Hopcroft et al.~\cite{hkks-tecl-04} were some of the first authors who
clustered snapshots with the help of a static clustering algorithm and
then tracked the found clusters over time.  They overcome the problem that even
small perturbations in the underlying graph may lead to significant
changes in the structure of the  found clusters in consecutive snapshots by
applying a greedy agglomerative approach several times with different
orderings of vertices.  Only clusters that remain stable under such
multiple clustering runs (so called \emph{natural} or \emph{consensus
  clusters}) are then considered for the tracking.

Another way to overcome the problem of unstable clusterings is to
explicitly incorporate \emph{temporal smoothness} in the clustering
process.  A first attempt in this direction was done by Chakrabarti et
al.~\cite{ckt-ec-06} in 2006; however, they clustered attributed data instead of
graphs.  Their idea is to exploit the knowledge about the previously
found clustering to find a clustering for the current time step that
is similar to the previous clustering (i.e., has low \emph{history
  cost}) and is still a good clustering also for the data in the
current time step (i.e., has high \emph{snapshot quality}).  Depending
on the clustering algorithm in which the temporal smoothness is
incorporated, this may lead to an objective function that needs to be
optimized or to an adaption of the input data or certain parameters.
Chakrabarti et al.\ examine two widely used clustering algorithms
within their framework; k-means and agglomerative hierarchical
clustering.  Their technique of incorporating temporal smoothness into
static clustering approaches has been established under the name
\emph{evolutionary clustering}.
%
It has been adapted to graph clustering by Kim and Han~\cite{kh-a-09},
Chi et al.~\cite{cszht-e-07}, G\"{o}rke et al.~\cite{gmssw-dgccm-11a}
and Xu et al.~\cite{xki-tcdsn-11}.  The \algo{FacetNet} approach by
Lin et al.~\cite{lczst-a-09} is based on a generic model that is
equivalent to the framework of Chakrabarti et al.\ under certain
assumptions.  The corresponding algorithms are described in more
detail in Section~\ref{sec:IncrClus}.  Note that the term
\emph{evolutionary} is not limited to evolutionary clustering (as
introduced by Chakrabarti et al.).  It is also used in many other
contexts, like, for example, in the context of evolutionary search
heuristics.

Under the term~\emph{dynamic graph clustering} we subsume the
remaining clustering approaches presented in this survey.  The
difference to evolutionary clustering, where in each time step a
static algorithm is applied from scratch, is that dynamic approaches
update existing information from previous time steps without
recalculating the whole clustering from scratch.  This can be done,
for example, by reusing parts of the previous clustering and just
updating local areas where the clustering has become
infeasible~\cite{arb-r-12} or, in case of greedy agglomerative
algorithms, initializing the current clustering with the previously
found clusters~\cite{trz-ilcid-13}.  Other approaches update auxiliary
information like sets of dense subsets~\cite{asks-d-12}, lists of well
connected neighbors~\cite{lwy-dcdwg-13},
eigenvectors~\cite{nxcgh-iscwa-07} or structures like cut
trees~\cite{ghw-dcc-11} or clique graphs~\cite{dyll-ik-12}.  Quite a
few approaches also combine these techniques.  Although most of these
dynamic approaches aim at reducing the running time, in many cases
updating previous information implicitly also leads to temporal
smoothness.  Other update methods admit the detection of cluster
events like splitting or merging~\cite{fbs-dengr-07,ghw-dcc-11}.

\paragraph{Evaluation of Clustering Methods.}
As the (rather fuzzy) definition of graph clustering according to the
intracluster density versus intercluster sparsity paradigm does not
correspond to a well defined optimization problem, the comparison of
different clustering approaches is inherently difficult.  One aspect
that should always be taken into account is scalability, i.e., to what
extent the running time increases with the graph size.  Disregarding
the influence of more or less efficient implementations and hardware
environments, the scalability of algorithms can be easily compared and
evaluated.  In some cases, it is possible to consider the immediate
usefulness of a clustering for a certain application, which allows to
compare different clusterings in a precise and well motivated way.  An
example for this is the use of dynamic graph clustering in the context
of routing protocols in Mobile Ad hoc Networks (MANETS), where
clusterings can be evaluated based on their impact on statistics as
Delivery Ratio or Average Delivery Time~\cite{dyt-t-09,ndyt-a-11}.
However, most applications do not yield such statistics, which is why
most authors focus on two main approaches to evaluate clusterings,
both of which are not undisputed.

The first one is the evaluation of clusterings with the help of a
\emph{ground truth} clustering. 
For real world data, in most cases this corresponds to
additional metadata
that indicate well motivated communities.  
In the context of
synthetic data this usually refers to clusters
\emph{implanted} in the graph structure during generation.  Usually, a
ground truth clustering either corresponds to a partition of the
objects in the context of algorithms finding complete clusterings, or
a set of subsets of objects in the context of algorithms admitting
overlapping clusters.  Ground truth clusterings can now be compared to
the outcome of an algorithm with the help of a suitable \emph{distance
  measure} on partitions or sets of objects.  In the next section we
introduce some distance measures that can be used to evaluate
clusterings based on ground truth clusterings.  Moreover,
Section~\ref{sec:benchAndGen} gives an overview of real world datasets
used in the literature both with and without metadata, and describes
some models that can be used to generate synthetic data.  In case the
metadata about real world data is not available in form of a ground
truth clustering, it is further possible to manually look into the
data and perform plausibility checks of the clusterings obtained.
However, this approach requires a lot of interpretation and, due to
the large size of some datasets, is often necessarily limited to a
subset of the data at hand.

The second main approach is the use of quality measures to evaluate
the goodness of a given clustering.  Using quality measures simplifies
the evaluation a lot, as it turns the inherently vague definition of
clustering into an explicit optimization problem.  Some algorithms use
these objective functions explicitly and optimize them by using for
example local search techniques or trying to find provably optimal
solutions
efficiently~\cite{bbp-fcddc-11,dnt-aaaac-13,dsttz-agami-10,dyt-t-09,gmssw-dgccm-11a,ndyt-a-11,rb-mcmms-13}.
Others use objective functions as an additional evaluation criterion,
although the algorithm itself does not explicitly optimize any
measure~\cite{gm-addhc-10,xcs-lrtic-13}; often, the authors then do
not claim to obtain the best values according to the considered
measure, but use the results as an additional sanity check to motivate
their approach, together with experiments involving ground truth
clusterings.  We will give the definitions of some commonly used
quality measures in the next section.

  For a further discussion on
the difficulties of evaluating community detection methods and a brief
history of method evaluation, see~\cite{lc-b-13}.

\subsection{Quality and Distance Measures for
  Clusterings}\label{subsec:measures}
In this section, we will give a short overview of quality measures assessing
the goodness of clusterings, followed by a discussion on what has to
be additionally considered in the dynamic scenario, and an introduction
to some frequently used distance measures that can be used to evaluate
the similarity of two clusterings.  To give a comprehensive overview
of all quality and distance measures used in the literature is beyond
the scope of this article; further information can be found for
example in the articles of Fortunato~\cite{f-c-09} and Wagner
et al.~\cite{ww-ccao-07}.

\paragraph{Quality Measures in Static Graphs.}
We will describe two main approaches to measure the quality of a
clustering in the static scenario, the first one is based on
\emph{balanced cuts} and the second on
\emph{null models}.  For better readability, we consider only
unweighted graphs in this section.  Note that all measures described
here can be generalized to weighted graphs in a straightforward way.

The use of cuts as a means to analyze community structures in networks
has a long tradition~\cite{z-ifmcf-77}.  A trivial community detection
algorithm could for example determine the minimum cut in a graph,
split the graph according to this cut, and recurse this procedure on
the resulting communities until some termination criterion, as for
example a desired number of clusters, is met.  This procedure at least
guarantees that two found communities or clusters are locally not too
strongly connected by intercluster edges.  Nevertheless, this often
leads to clusters of very uneven sizes; especially in the presence of
low degree vertices, minimum cuts tend to separate only one or few
vertices from the remainder of the graph.  For this reason, clustering
algorithms typically build upon \emph{balanced cuts}, i.e., cuts that
simultaneously cross few edges and split the graph in two
approximately equal sized parts.

Probably the first formal definitions of balanced cuts used in the
context of graph clustering are the measure \emph{conductance} and
\emph{expansion}~\cite{kvv-ocgbs-04}.  For a subset $S$, let $e(S, V
\setminus S)$ denote the number of edges linking vertices in $S$ with
vertices in $V \setminus S$.  Furthermore, the \emph{volume}
$\operatorname{vol}(S) := \sum_{v \in S} \operatorname{deg}(v)$ of a
subset of vertices $S$ is defined as the sum of the degrees of its
vertices.  Then, the conductance $\operatorname{cond}$ of a cut $(S, V
\setminus S)$ can be written as:
\ifreview
\begin{linenomath}
\fi
\begin{equation*}
  \operatorname{cond}(S, V \setminus S) = \frac{e(S, V \setminus S)}{\min\{ \operatorname{vol}(S), \operatorname{vol}(V \setminus S) \}} 
\end{equation*}
\ifreview
\end{linenomath}
\fi
Many variations thereof exist, most of which either replace the volume
by other notions of cluster size, or use a slightly different
tradeoff between the cut size and the sizes of the two induced parts.
We give the definition of two of these variation, namely
\emph{expansion} and \emph{normalized cut}:
\ifreview
\begin{linenomath}
\fi
\begin{align*}
  \operatorname{exp}(S, V \setminus S) &= \frac{e(S, V \setminus S)}{\min\{ |S|, |V \setminus S| \}} \\
  \operatorname{ncut}(S, V \setminus S) &= \frac{e(S, V \setminus
    S)}{\operatorname{vol}(S)} + \frac{e(S, V \setminus
    S)}{\operatorname{vol}(V \setminus S)}
\end{align*}
\ifreview
\end{linenomath}
\fi
The latter definition is especially popular in the field of image
segmentation~\cite{sm-ncis-00}.
Finding
an optimal cut with respect to any of the three definitions above is
$\mathcal{NP}$-hard~\cite{lr-m-99,sm-ncis-00,us-onpcs-06}, which is
why divisive algorithms are usually based on approximation algorithms
or heuristics.
It remains to mention that cut-based measures are closely
related to spectral clustering techniques~\cite{l-atsc-07}.

It is not immediately clear how the above measures can be used to
evaluate whole clusterings.  One possibility is to associate two
values with each cluster $C$, one that evaluates the cut that
separates the cluster from the rest of the graph and another
evaluating all cuts within the subgraph that is induced by~$C$.  In
the context of conductance, this leads to the following definition of
\emph{inter-} and \emph{intracluster conductance} of a
cluster~$C$~\cite{bgw-egcme-07}:
\ifreview
\begin{linenomath}
\fi
\begin{align*}
 \operatorname{\text{intercluster conductance}}(C) &= \operatorname{cond}(C, V \setminus C)\\
 \operatorname{\text{intracluster conductance}}(C) &= \min_{S \subset C} \{\operatorname{cond}(S, C \setminus S)\}
\end{align*}
\ifreview
\end{linenomath}
\fi
In a good clustering according to the intracluster density versus
intercluster sparsity paradigm, the intracluster conductance of the
clusters is supposed to be high while their intercluster conductance
should be low.  An overall value for the intracluster conductance of a
whole clustering can then be obtained by taking, for example, the
minimum or average of the {intra\-cluster} conductance over all
clusters~\cite{gsw-dcgc-11b}.  Analogously, the intercluster
conductance of a clustering can be defined as the maximum intercluster
conductance over all clusters.  This leads to a bicriterial
optimization problem.  Calculating the intracluster conductance of a
cluster is $\mathcal{NP}$-hard, which immediately follows from the
$\mathcal{NP}$-hardness of finding a cut with optimum conductance.
The same holds if we replace conductance by expansion or normalized
cut.  Hence, most clustering approaches that aim at solving the
resulting optimization problem are again based on approximation
algorithms or heuristics.  The cut clustering algorithm of Flake et
al.~\cite{ftt-gcmct-04,ghw-dcc-11}, however, guarantees at least a
lower bound of the intracluster expansion of the found clusters.
%
In principal, these cut based criteria can be also used to evaluate
overlapping clusterings, although they are much more
common in the context of complete clusterings.

Another approach to measure the goodness of clusterings that has
gained a lot of attention during the last decade is the use of null
models.  Roughly speaking, the idea behind this is to compare the
number of edges within clusters to the expected number of edges in the
same partition, if edges are randomly rewired.  The most popular
measure in this context is the \emph{modularity} of a clustering as
defined by Girvan and Newman~\cite{ng-fecsn-04} in
2004.  Let $e(C)$
denote the number of edges between the vertices in cluster $C$.  Then,
the modularity $\operatorname{mod}(\mathcal{C})$ of a (complete)
clustering~$\mathcal{C}$ can be defined as
\ifreview
\begin{linenomath}
\fi
\begin{equation*}
 \operatorname{mod}(\mathcal{C}) = \sum_{C \in \mathcal{C}} \frac{e(C)}{m} - \sum_{C \in \mathcal{C}} \frac{{\operatorname{vol}(C)}^2}{4m^2}.
\end{equation*}
\ifreview
\end{linenomath}
\fi
Here, the first term measures the actual fraction of edges within
clusters and the second the expectation of this value after random
rewiring, given that the probability that a rewired edge is incident
to a particular vertex is proportional to the degree of this vertex in
the original graph.  The larger the difference between these terms,
the better the clustering is adjusted to the graph structure.
The corresponding optimization problem is $\mathcal{NP}$-hard~\cite{bdgghnw-omc-08}.
Modularity can be generalized to weighted~\cite{n-awn-04} and
directed~\cite{adfg-s-07,ln-csdn-08} graphs, to overlapping or fuzzy
clusterings~\cite{nmcm-edmdg-09,scch-d-09}, and to a local scenario,
where the goal is to evaluate single
clusters~\cite{czg-dclni-09,c-f-05,lwp-elcsl-06}.  Part of its
popularity stems from the existence of heuristic algorithms that
optimize modularity and that are able to cluster very large graphs in
short time~\cite{bgll-f-08,og-a-13,rn-m-11}.  In
Section~\ref{sec:IncrClus}, we will describe some generalizations of
these algorithms to the dynamic setting.  Furthermore, in contrast to
many other measures and definitions, modularity does not depend on any
parameters.  This might explain why it is still widely used, despite
some recent criticism~\cite{bf-rlcd-07}.

\paragraph{Quality Measures in Evolving Graphs.}
In the context of dynamic graph clustering, we aim at clusterings of
high quality for each snapshot graph.  Compared to the static
approach, as discussed in Section~\ref{sec:introduction},
temporal smoothness becomes an additional dimension.  Speaking in
terms of objective functions, we would like to simultaneously optimize
the two criteria quality and temporal smoothness.

As already mentioned before,
one approach to obtain this is introduced by Chakrabarti et
al.~\cite{ckt-ec-06}.  The idea is to measure the \emph{snapshot
  quality}~sq of the current clustering~$\mathcal{C}_t$ at time
step~$t$ (with respect to the current snapshot~$\mathcal{G}_t$) by a
(static) measure for the goodness of a clustering.  Similarly, the
smoothness is measured by the \emph{history cost}~$\text{hc}$ of the
current clustering, which is usually defined as the distance of the
current clustering to the previous clustering~$\mathcal{C}_{t-1}$ at
time step~$t-1$.
%
The snapshot quality could for example be measured by modularity and
the smoothness by any of the distance measures introduced in the next
paragraph.  The goal is then to optimize a
linear combination of both measures, where $\alpha$ is an input
parameter that determines the tradeoff between quality and
smoothness:
\ifreview
\begin{linenomath}
\fi
\begin{equation*}
  \text{minimize } \alpha \cdot \text{sq}(\mathcal{C}_t, \mathcal{G}_t) - (1-\alpha) \cdot
  \text{hc}(\mathcal{C}_t, \mathcal{C}_{t-1}).
\end{equation*}
\ifreview
\end{linenomath}
\fi

Closely related to this approach, but not relying on an explicit
distance measure, is the claim that a good clustering of the snapshot
at time step~$t$ should also be a good clustering for the snapshot at
time step~$t-1$.  This is based on the underlying assumption that
fundamental structural changes are rare.  Hence,
linearly combining the snapshot quality of the current clustering with
respect to the current snapshot~$\mathcal{G}_t$ and the previous
snapshot~$\mathcal{G}_{t-1}$ yields a dynamic quality measure, which
can be build from any static quality measure:
\ifreview
\begin{linenomath}
\fi
\begin{equation*}
  \text{minimize } \alpha \cdot \text{sq}(\mathcal{C}_t,\mathcal{G}_t) + (1-\alpha) \text{sq}(\mathcal{C}_t, \mathcal{G}_{t-1}).
\end{equation*}
\ifreview
\end{linenomath}
\fi
This causes the clustering at time step~$t$ to also take the structure
of snapshot $\mathcal{G}_{t-1}$ into account, which implicitly enforces
smoothness.  Takaffoli et al.~\cite{trz-ilcid-13} apply this approach
in the context of modularity, and Chi et al.~\cite{cszht-e-07} in the
context of spectral clustering; both will be discussed in
Section~\ref{sec:IncrClus}.

\paragraph{Distance Measures for Clusterings.}
In the context of graph clustering, distance measures have three main
applications.  First, similar to static clustering, they can be used
to measure the similarity to a given ground truth clustering.  Second,
they can be used as a measure of smoothness, for example by comparing
the clusterings of adjacent time steps.  Third, they are useful in the
context of event detection; a large distance between two consecutive
clusterings may indicate an event.  A plethora of different measures
exist in the literature, none of which is universally accepted.  For
this reason, we will only introduce the measures used by the dynamic
algorithms we describe in Section~\ref{sec:IncrClus}.  This includes
the probably best known index in the context of clustering, the
\emph{normalized mutual information}.  If not mentioned otherwise, all
clusterings considered in this section are assumed to be complete.

Mutual information has its roots in information theory and
is based on the notion of the entropy of a clustering~$\mathcal{C}$.
For a cluster~$C\in \mathcal{C}$, let $P(C) := |C|/n$.  
With that, the entropy~$\mathcal{H}$ of $\mathcal{C}$ can be defined
as
\ifreview
\begin{linenomath}
\fi
\begin{equation*}
 \mathcal{H}(\mathcal{C}) := - \sum_{C \in \mathcal{C}} P(C) \log_2 P(C)
\end{equation*}
\ifreview
\end{linenomath}
\fi
Similarly, given a second clustering~$\mathcal{D}$, with~$P(C,D) := |C \cap D|/n$, the conditional
entropy~$H(\mathcal{C} | \mathcal{D})$ is defined as
\ifreview
\begin{linenomath}
\fi
\begin{equation*}
  \mathcal{H}(\mathcal{C} | \mathcal{D}) := \sum_{C \in \mathcal{C}} \sum_{D \in \mathcal{D}} P(C,D) \log_2 \frac{P(C)}{P(C,D)}
\end{equation*}
\ifreview
\end{linenomath}
\fi
Now the \emph{mutual information}~$\mathcal{I}$
of~$\mathcal{C}$ and~$\mathcal{D}$ can be defined as
\ifreview
\begin{linenomath}
\fi
\begin{equation*}
  \mathcal{I}(\mathcal{C}, \mathcal{D}) := \mathcal{H}(\mathcal{C}) - \mathcal{H}(\mathcal{C} | \mathcal{D}) = \mathcal{H}(\mathcal{D}) - \mathcal{H}(\mathcal{D} | \mathcal{C}) = \sum_{C \in \mathcal{C}} \sum_{D \in \mathcal{D}} P(C,D) \log_2 \frac{P(C,D)}{P(C) P(D)}
\end{equation*}
\ifreview
\end{linenomath}
\fi
Informally, this is a measure of how much information the knowledge that a
vertex belongs to a certain cluster in clustering $\mathcal{D}$ yields
about its cluster id in $\mathcal{C}$.  Several normalizations of this
measure exist; according to Fortunato~\cite{f-c-09}, the most commonly
used normalization is the following notion of \emph{normalized mutual
  information (NMI)}, which maps the mutual
information to the interval~$[0,1]$:
\ifreview
\begin{linenomath}
\fi
\begin{equation*}
  \text{NMI}(\mathcal{C}, \mathcal{D}) = \frac{2 \mathcal{I}(\mathcal{C}, \mathcal{D})}{\mathcal{H}(\mathcal{C}) + \mathcal{H}(\mathcal{D})}
\end{equation*}
\ifreview
\end{linenomath}
\fi
Technically, this is not a distance but a similarity measure, as high
values of NMI indicate high correlation between the clustering.  If
need be, it can be easily turned into a distance measure by
considering~$1-\text{NMI}(\mathcal{C}, \mathcal{D})$.  
There also exists a generalization to overlapping clusterings~\cite{lfk-d-09}.
Among the
approaches we describe in Section~\ref{sec:IncrClus}, Yang et
al.~\cite{yczj-db-11}, Cazabet et al.~\cite{cah-d-10} and Kim and Han~\cite{kh-a-09} use mutual
information to compare against ground truth clusterings. In contrast
to that, Lin et al.~\cite{lczst-a-09} use it to compare the time step
clusterings to the communities of the aggregated graph, which can be
seen as both a measure of smoothness and comparison to some kind of
ground truth clustering.  Wang et al.~\cite{lwy-dcdwg-13} use NMI both to measure
the similarity of a clustering to a generated ground truth clustering and to
compare the results of an approximation algorithm to clusterings found
by an exact algorithm (according to their definition of clusters).

Aynaud and Guillaume~\cite{ag-s-10} use, as an alternative to NMI, the
minimum number of vertex moves necessary to convert one clustering
into the other as a measure of distance.  Their main argument to
 consider this approach is
that absolute values are far easier to interpret.

Another very intuitive measure for the distance between two partitions
is the \emph{Rand index} introduced by Rand~\cite{r-ocecm-71} in
1971.  Let~$s(\mathcal{C},\mathcal{D})$ be the number of vertex
pairs that share a cluster both in~$\mathcal{C}$ and~$\mathcal{D}$
and~$d(\mathcal{C}, \mathcal{D})$ the number of vertex pairs that are
in different clusters both in~$\mathcal{C}$ and~$\mathcal{D}$.  With
that, the Rand index $\mathcal{R}$ of $\mathcal{C}$ and $\mathcal{D}$
is defined as
\ifreview
\begin{linenomath}
\fi
\begin{equation*}
 \mathcal{R}(\mathcal{C}, \mathcal{D}) := 1-\frac{s(\mathcal{C}, \mathcal{D}) + d(\mathcal{C}, \mathcal{D})}{\binom{n}{2}}
\end{equation*}
\ifreview
\end{linenomath}
\fi
This corresponds to counting the number of vertex pairs where both
clusterings disagree in their classification as intracluster or
intercluster pair, followed by a normalization.  Delling et
al.~\cite{dggw-ecgc-08} argue that this measure is not appropriate in
the context of graph clustering, as it does not consider the topology
of the underlying graph.  They propose to only consider vertex pairs
connected by an edge, which leads to the \emph{graph based} Rand
index.  This graph based version is used by G\"{o}rke et
al~\cite{gmssw-dgccm-11a} to measure the distance between clusterings
at adjacent time steps.

Chi et al.~\cite{cszht-e-07} use the \emph{chi square statistic} to
enforce and measure the similarity between adjacent clusterings.  The
chi square statistic was suggested by Pearson~\cite{p-o-00} in
1900 to test for independence in a bivariate distribution.  In the
context of comparing partitions, different variants
exist~\cite{m-ewlcs-01}; the version used by Chi et al.~is the
following:
\ifreview
\begin{linenomath}
\fi
\begin{equation*}
 \chi^2(\mathcal{C}, \mathcal{D}) = n \cdot \left(\sum_{C \in \mathcal{C}} \sum_{D \in \mathcal{D}} \frac{|C \cap D|}{|C| \cdot|D|} -1\right)
\end{equation*}
\ifreview
\end{linenomath}
\fi

\subsection{Applications}\label{subsec:app}
Graph clustering has possible applications in many different
disciplines, including biology and sociology. Biologists are for
example interested in how diseases spread over different communities,
sociologists often focus on cultural and information transmission.
Many of the networks analyzed in these areas have a temporal dimension
that is often neglected; taking it into account potentially increases
the usefulness of clustering for the respective application and at the
same time evokes new challenges like for example the involvement of
temporal smoothness.  In the context of social networks, the benefit
of temporal smoothness becomes in particular obvious, since social
relations and resulting community structures are not expected to
change frequently.  Giving an exhaustive list of application areas is
beyond the scope of this article; some further information can be
found in the overview article of Fortunato~\cite{f-c-09}.  Instead, we
will give some examples where clustering approaches designed for
evolving graphs have clearly motivated advantages over static
approaches.

A little-known but very interesting application of graph clustering is
the use in graph drawing or visualization algorithms.  The general
idea is to first cluster the vertices of the graph and then use this
information in the layouting steps by placing vertices in the same
community in proximity of each other.  This has several advantages:
The layout makes the community structure of the graph visible, which
is desirable in many applications.  Furthermore, he intracluster
density versus intracluster sparsity paradigm causes many edges to be
within clusters, which in turn corresponds to small edge lengths.
Last but not least, layout algorithms that use clustering as a
preprocessing step are usually quite fast.  As an example, Muelder and
Ma have used clustering algorithms in combination with layouts based
on treemaps~\cite{mm-atbmr-08} and space filling
curves~\cite{mm-rglus-08}.  A straightforward extension to these
approaches is the task to visualize dynamic
graphs~\cite{smm-cvnld-13}.  Dynamic clustering algorithms can help in
this context to reduce the running time for the preprocessing in each
time step.  Furthermore, if they are additionally targeted at
producing smooth clusterings, this results in smoother layouts, or, in
terms of layout algorithms, in a good preservation of the mental map.

  Another interesting application of dynamic graph clustering is its
  use in routing protocols in Mobile Ad hoc Networks (MANETS).
  \emph{On-Demand} forwarding schemes for this problem discover paths
  in the network only when receiving concrete message delivery
  requests.  It has been shown that ``routing strategies based on the
  discovery of modular structure have provided significant performance
  enhancement compared to traditional schemes''.  \cite{dyt-t-09} Due
  to the mobility of actors in the network, the resulting topology is
  inherently dynamic; recomputing the clustering whenever a change
  occurs is costly and requires global information.  This motivated a
  number of online algorithms for modularity based dynamic clustering
  algorithms, with experiments showing that the use of the dynamic
  clustering improved the performance of forwarding schemes in this
  scenario~\cite{dyt-t-09,ndyt-a-11}.  Another interesting aspect of
  this application is that ``consistent modular structures with
  minimum changes in the routing tables''~\cite{dyt-t-09} are
  desirable, again motivating temporal smoothness.

\section{Online Graph Clustering Approaches}\label{sec:IncrClus}
In this section we introduce current clustering algorithms and
community detection approaches for evolving graphs.  As discussed
above, we consider only algorithms that operate in an online scenario,
i.e., that do not use information from future time steps, and are
incremental in the sense that they incorporate historic information
from previous time steps to achieve temporal smoothness or a better
running time.  We use different categories to classify the approaches
presented here.  Some categories are associated with particular
algorithmic techniques, other categories with applications or the form
of the resulting clusterings. 
Apart from these categories, the \algo{GraphScope}
approach~\cite{sypf-gspfm-07} is presented at the beginning of this
section, as it is one of the first and most cited dynamic approaches.
The section concludes with two further approaches, which do not fit
into one of the previous categories.  

\paragraph{GraphScope.}
The \algo{GraphScope} approach by Sun et al.~\cite{sypf-gspfm-07} is one of
the first and most cited dynamic clustering approaches so far.
However, contrary to the notion of communities as densely connected
subgraphs, \algo{GraphScope} follows the idea of block modeling, which is
another common technique in sociology.  The aim is to group actors in
social networks by their role, i.e., structural equivalence. Two
actors are equivalent if they interact in the same way with the same
actors (not necessarily with each other). This is, the subgraph
induced by such a group may be disconnected or even consisting of an
independent set of vertices.  The latter is the case in approaches
like \algo{GraphScope} that consider bipartite graphs of source and
destination vertices and seek for groups of equivalent vertices in
each part, i.e., groups consisting either of source or destination
vertices.  Furthermore, instead of independent snapshots, \algo{GraphScope}
considers whole graph segments, which are sequences of similar
consecutive snapshots that (w.l.o.g.)  have all the same number of
sources and destinations.
%
The main idea is the following.  Given a graph segment and a partition
of the vertices in each part (the same partition for all snapshots in
the graph segment), the more similar the vertices are per group the
cheaper are the encoding costs for the graph segment
using an appropriate encoding scheme based on a form of Minimum
Description Length (MDL)~\cite{r-m-78}.  This is, \algo{GraphScope} seeks for
two partitions, one for each part of the bipartite input graph, that
minimize the encoding costs with respect to the current graph segment.
It computes good partitions in that sense by an iterative greedy
approach.  Based on the same idea, the MDL is further used to decide
whether a newly arriving snapshot belongs to the current graph segment
or starts a new segment.  If the new snapshot belongs to the current
graph segment, the two partitions for the graph segment are updated
starting the iteration with the previous partitions.  If the new
snapshot differs too much from the previous snapshots, a new segment
is started.  In order to find new partitions in the new segment, the
iterative greedy approach is either initialized with the partitions of the
previous graph segment or the iterations are done from scratch.
The latter can be seen as a static version of \algo{GraphScope}.
An experimental comparison on real world data proves a
much better running time of the dynamic approach with respect to the
static approach.  Additional experiments further illustrate that the
found source and destination partitions correspond to semantically
meaningful clusters.  Although this approach focuses on bipartite
graphs, it can be easily modified to deal with unipartite graphs, by
constraining the source partitions to be the same as the destination
partitions~\cite{c-appfg-04}.

\paragraph{Detecting Overlapping Dense Subgraphs in Microblog
  Streams.}
The approaches of Angel at al.~\cite{asks-d-12} and Agarwal et
al.~\cite{arb-r-12} both date back to the year 2012 and aim at
real-time discovery of emerging events in microblog streams, as
provided for example by Twitter.  To this end, they model the
microblog stream as an evolving graph that represents the correlation
of keywords occurring in the blogs or messages.  In this keyword
graph, they seek for groups of highly correlated keywords, which
represent events and are updated over the time.  Since a keyword may
be involved in several events, these groups are allowed to overlap.
The main differences between both attempts is the definition of the
correlation between keywords and the definition of the desired
subgraphs.  Angel et al.\ consider two keywords as correlated if they
appear together in the same message.  Two keywords are the stronger
correlated the more messages contain them together.  The messages are
considered as a stream and older messages time out. This results in
atomic updates of the keyword graph.  In contrast, Agarwal et al.\
consider multiple changes in the keyword graph resulting from a
sliding time window. They consider two keywords as correlated if they
appear in (possibly distinct) messages of the same user within the
time window.  Furthermore, they ignore all edges representing a
correlation below a given threshold, which results in an unweighted
keyword graph.

Regarding the group detection, 
%
%
Angel et al.\ introduce an algorithm called \algo{DynDens} that
considers a parameterized definition of density, which covers most
standard density notions.  Based on this definition, a set is dense if
its density is greater than a given threshold.  In order to return all
dense subgraphs for each time step, a set of almost dense subgraphs is
maintained over time that has the property that after a change in the
keyword graph each (possibly new) dense subgraph contains one of the
maintained almost dense subgraphs.  Hence, the almost dense subgraphs
can be iteratively grown to proper density, thus finding all new dense
subgraphs after the change.  With the help of an appropriate data
structure the almost dense subgraphs can be maintained efficiently
with respect to time and space requirements.
In order to give experimental evidence of the feasibility of their
approach, the authors have built a live demo for their techniques on
Twitter-tweets and
provide, besides a qualitative evaluation, 
a comparison with a simple static baseline approach that periodically
recomputes all dense subgraphs.
This static approach took that much time
that it was able to compute the set of new events only every 48 to 96
minutes, compared to a real time event identification performed by the
dynamic approach.
%
Instead of dense subgraphs, Agarwal et al.\ seek for subgraphs that
possess the property that each edge in the subgraph is part of a cycle
of length at most~4.  This property is highly local, and thus, can be
updated efficiently.  An experimental study on real-world data
comparing the dynamic approach to a static algorithm that computes
biconnected subgraphs confirms the efficiency of the local
updates.

\paragraph{Other Approaches Admitting the Detection of Overlapping
  Clusters.}
While two of the following approaches indeed return overlapping
clusters, the remaining approaches use a cluster definition that
basically admits overlapping clusters, but the algorithmic steps for
finding these clusters are designed such that they explicitly avoid
overlaps.  The most common reason for such an avoidance is the fact
that tracking overlapping clusters over time is even more difficult
than tracking disjoint clusters, which is why most of the existing
tracking frameworks require disjoint clusters.

Takaffoli et al.~\cite{trz-ilcid-13} incrementally apply a method
inspired by Chen et al.~\cite{czg-dclni-09} that basically returns
overlapping clusters. In order to apply, in a second step, an
independent event detection framework~\cite{tfsz-t-11} that requires
disjoint clusters, they however rework this method such that it
prevents overlaps.
The idea is to greedily grow clusters around core sets that serve as
seeds.  In doing so the aim is to maximize the ratio of the average
internal degree and the average external degree of the vertices in the
cluster, only considering vertices with a positive internal and
external degree, respectively.  The reworking step then allows a
vertex to also leave its initial core set, which admits the shrinking
of clusters and the prevention of overlapping clusters.  For the first
snapshot in a dynamic scenario the initial core sets are single
vertices (static version), whereas any further snapshot is clustered
using the clusters of the previous snapshot as initial core sets
(dynamic approach).  If a previous cluster decomposes into several
connected components in the current snapshot, the authors consider
each of the connected components as a seed.  Compared to the static
version applied to each snapshot independently and also compared to
the \algo{FacetNet} approach~\cite{lczst-a-09} (which we introduce in
the category of generative models), at least for the Enron network
tested by the authors, the dynamic attempt results in a higher average
community size and a higher \emph{dynamic modularity} per snapshot.
The latter is a linear combination of the modularity values of the
current clustering with respect to the current snapshot and the
previous snapshot (see also the quality measures in evolving graphs
presented in Section~\ref{subsec:measures}).

Kim and Han~\cite{kh-a-09} present an evolutionary clustering method,
which incorporates temporal smoothness to \algo{Scan}~\cite{xyfa-scana-07}, a
popular adaption of the (static) density-based data clustering
approach \algo{dbScan}~\cite{eksx-adbad-96} to graphs.
The new idea, compared to the idea of evolutionary clustering by
Chakrabarti et al.~\cite{ckt-ec-06}, is that instead of minimizing a
cost function that trades off the snapshot quality and the history
quality at every time step, the same effect can be achieved by
adapting the distance measure in \algo{Scan}.  As usual for
\algo{Scan}, the authors define an $\varepsilon$-neighborhood of a
vertex with respect to a distance measure, such that the resulting
$\varepsilon$-neighborhood consists of core vertices and border
vertices.  A cluster is then defined as the union of
$\varepsilon$-neighborhoods, each of which having size at
least~$\eta$,
that overlap in at least one core vertex.  This kind of clusters can
be easily found by a BFS in the graph.
This initially yields clusters that may overlap in some border
vertices. However, by ignoring vertices that are already assigned to a
cluster during the BFS, disjoint clusters can be easily enforced.  A
vertex that is not found to be a member of a cluster, is classified as
noise.  By iteratively adapting~$\varepsilon$ the authors additionally
seek for a clustering of high modularity.  When a good clustering is
found for the current snapshot, temporal smoothness is incorporated by
adapting the distance measure that characterizes the
$\varepsilon$-neighborhoods in the next time steps allowing for the
distance in the current snapshot and the not yet adapted distance in
the next snapshot.  Finally, the authors also propose a method for
mapping the clusters found in consecutive snapshots, based on mutual
information.
On synthetic networks of variable numbers of clusters
the proposed approach outperformed \algo{FacetNet}~\cite{lczst-a-09}
with respect to clustering accuracy and running time.

Another approach that is also based on \algo{dbScan} and is very similar to
\algo{Scan}, is called \algo{DenGraph} and is presented by Falkowski et
al.~\cite{fbs-dengr-07}.  In contrast to Kim and Han, Falkowski et
al.\ do not focus on an evolutionary clustering approach, but
introduce dynamic update techniques to construct a new clustering
after the change of a vertex or an edge in the underlying graph.
These updates are done by locally applying again an BFS (as for the
static \algo{Scan} approach), but just on the vertices that are close to the
change, thereby updating the cluster ids.  Experiments on the Enron
data set suggest that \algo{DenGraph} is quite fast but also relatively
strict as it reveals only small, very dense groups while many vertices
are categorized as noise. The dynamic \algo{DenGraph} version proposed
in~\cite{fbs-dengr-07} returns disjoint clusters, while
in~\cite{f-cacad-09} Falkowski presents a dynamic version for
overlapping clusters.

A dynamic algorithm that is not based on a static clustering algorithm
but also produces overlapping clusters is proposed by Cazabet et
al.~\cite{cah-d-10}.  In each time step, clusters are updated in the
following way.  First, it is determined if a new seed cluster, i.e., a
small clique of constant size has emerged due to edge updates.  Then,
existing clusters and seed clusters are extended by additional
vertices.  To that end, for each cluster~$C$, two characteristics are maintained.
The first of these characteristics corresponds to the average percentage of vertices a vertex in $C$ can reach in its cluster by a path of length $2$.
Very similar, the second characteristic corresponds to the average percentage of vertices a vertex in $C$ can reach in its cluster by at least two distinct path of length $2$.
A vertex that is not in $C$ may be included into $C$ if, roughly
speaking, this improves both of these characteristics.  In a last
step, clusters that share a certain percentage of vertices with
another cluster are discarded.  The goal of this approach is not
primarily to get good clusterings for each time step but to get a good
clustering of the last time step by taking the evolution of the
network into account.
Nevertheless, the approach per se is capable of clustering dynamic
networks in an online scenario.  Other overlapping approaches are
categorized according to a different focus and are thus described at
another point.  For an overview on all overlapping approaches
presented in this survey see Table~\ref{tab:properties_dyn}.

\paragraph{Algorithms Maintaining Auxiliary Structures.}
The following two approaches consider atomic changes in the given
graph and aim at efficiently updating structurally clearly defined
clusters, which are obtained by a simple operation on an auxiliary
structure.  In the first approach, by Duan et al.~\cite{dyll-ik-12},
the auxiliary structure is a graph that represents the overlap of
maximal cliques in the input graph, and the final clusters result from
the connected components of this graph. In the second approach, by
G\"orke et al.~\cite{ghw-dcc-11}, a partial cut tree (or Gomory-Hu
tree~\cite{gh-mtnf-61}) is maintained and the clusters result from the
subtrees obtained by deleting a designated vertex in this tree.  The
latter approach further incorporates possibly given edge weights of
the input graph.

The dynamic clique-clustering approach of Duan et
al.~\cite{dyll-ik-12} is a dynamic version of the clique percolation
method (\algo{PCM}) of Der{\'e}nyi et al.~\cite{dpv-cprn-05}, which is again
a special case of a
more general clique-clustering framework proposed by Everett and
Borgatti~\cite{eb-aco-98}.  The framework by Everett and Borgatti
applies an arbitrary clustering algorithm to a weighted auxiliary
graph~$H$ that represents the overlap of the maximal cliques in the
input graph.  In the special case considered by Der{\'e}nyi et al.\
and Duan et al., the auxiliary graph~$H$ just encodes if two maximal
cliques (of at least size~$k$) overlap in at least $k-1$ vertices, and
thus, is an unweighted graph.  More precisely,~$H$ is a graph where
the maximal cliques in the input graph represent the vertices and two
vertices are connected by an edge if and only if the corresponding
cliques share at least~$k-1$ vertices.  As clustering algorithm
Der{\'e}nyi et al.\ and Duan et al.\ simply choose a DFS, which
returns the connected components of~$H$, which induce overlapping
clusters in the original graph.
%
The running time of this approach is dominated by the computation of
maximal cliques, which is exponential in the number of
vertices~\cite{bk-facug-73}.
The proposed dynamic version is then straightforward.  For each type
of change, the authors give a procedure to update the auxiliary
graph~$H$ as well as the DFS-tree~$T$, which indicates the connected
components of~$H$.  In doing so, the insertion of an edge is the only
change where the computation of new maximal cliques becomes necessary
in parts of the input graph.  All other changes can be handled by
updating the overlap of previous cliques and adapting edges in~$H$
and~$T$.  Hence, the more changes in the dynamic input graph are
different from edge insertions, the better the dynamic approach
outperforms the static approach, which computes all cliques from
scratch after each change.

The dynamic cut-clustering approach of G\"orke et
al.~\cite{ghw-dmctc-09,ghw-dcc-11} is a dynamic version of the static
cut-clustering algorithm of Flake et al.~\cite{ftt-gcmct-04}, based on
updating a partial Gomory-Hu tree~\cite{gh-mtnf-61} of an
extended input graph~$G_\alpha$.  The graph~$G_\alpha$ is obtained
from the input graph~$G$ by inserting an artificial vertex~$q$ and
artificial edges, each having weight~$\alpha$, between~$q$ and each
vertex in~$G$.  The input parameter~$\alpha$ determines the coarseness
of the resulting clustering.
A Gomory-Hu tree~$T$ for~$G_\alpha$ then is a weighted tree on the
vertices of~$G_\alpha$ that represents a minimum $s$-$t$-cut for each
vertex pair in~$G_\alpha$.  More precisely, deleting an edge~$\{s,t\}$
in~$T$ decomposes~$T$ in two subtrees inducing a minimum $s$-$t$-cut
in the underlying graph.  The weight assigned to the deleted edge
in~$T$ further corresponds to the costs of the induced minimum
$s$-$t$-cut.  For two non-adjacent vertices $u$ and~$v$ in~$T$, the
minimum $u$-$v$-cut is given by a cheapest edge on the path from~$u$
to~$v$ in~$T$.  In order to obtain the final complete clustering, the artificial
vertex~$q$ is deleted from~$T$, resulting in a set of subtrees
inducing the clusters.  Due to the special properties of the minimum
$s$-$q$-cuts that separate the resulting clusters, Flake et al.\ are
able to prove a guarantee (depending on~$\alpha$) of the intercluster
expansion and the intracluster expansion of the resulting clustering,
which in general is NP-hard to compute (cp.~the cut-based quality
measures introduced in Section~\ref{subsec:measures}).
The dynamic version of the cut-clustering algorithm determines which
parts of the current Gomory-Hu tree of~$G_\alpha$ become invalid due
to an atomic change in~$G$ and describes how to update these parts
depending on the type of the atomic change.  The result is a cut
clustering of the current graph~$G$ with respect to the same parameter
value~$\alpha$ as in the previous time step.  The most difficult and
also (in theory) most time consuming type of an update is the update
after an edge deletion.  However, in most real world instances the
actual effort for this operation is still low, as shown by an
experimental evaluation on real world data.  
We stress that there also exists another
attempt~\cite{sm-dagcu-06,sm-dagcu-07} that claims to be a dynamic
version of the cut clustering algorithm of Flake et al., however,
G\"orke et al.\ showed that this attempt is erroneous beyond
straightforward correction.  Doll et al.~\cite{dhw-dhmctc-11} further
propose a dynamic version of the hierarchical cut-clustering algorithm
that results from varying the parameter value~$\alpha$, as shown by
Flake et al.~\cite{ftt-gcmct-04}.

\paragraph{Spectral Graph Clustering Methods.}
The main idea of static spectral graph clustering is to find an
$r$-dimensional placement of
the vertices such that vertices that form a cluster in an appropriate
clustering with respect to a given objective are close to each other
while vertices that are assigned to different clusters are further
away from each other.  This can be done by considering the spectrum of
a variation of the adjacency matrix, like for example the Laplacian matrix in the context of the normalized cut objective~\cite{l-atsc-07}.  More precisely, many desirable objectives result in
optimization problems that are solved by the eigenvectors associated
with the top-$r$ eigenvalues of a variation of the adjacency matrix
that represents the objective.  The rows of the $n$-by-$r$ matrix
formed by these eigenvectors then represent $r$-dimensional
coordinates of the vertices that favor the objective.  The final
clustering is then obtained by applying, for example, $k$-means to these data
points.

The \algo{EvolSpec} algorithm by Chi et al.~\cite{cszht-e-07} conveys
this concept to a dynamic scenario by introducing objectives that
incorporate temporal smoothness.  Inspired by Chakrabarti et
al.~\cite{ckt-ec-06}, the authors linearly combine snapshot costs and
temporal costs of a clustering at time step~$t$, where the temporal
costs either describe how well the current clustering clusters
historic data in time step~$t-1$ or how different the clusterings in
time step~$t$ and~$t-1$ are.  For both quality
measures, they give the matrices that
represent the corresponding objectives, and thus, allow the use of
these measures in the context of spectral graph clustering.

Ning et al.~\cite{nxcgh-i-10} show how to efficiently update the
eigenvalues and the associated eigenvectors for established objectives
if an edge or a vertex in the underlying graph changes.  Compared to
the static spectral clustering, which takes~$O(n^{3/2})$ time, this
linear incremental approach saves a factor of~$n^{1/2}$.  An
experimental evaluation of the running times on Web-blog data
(collected by the NEC laboratories) confirm this theoretical result.
The fact that the updates yield only approximations of the desired
values is not an issue, as further experiments on the approximation
error and an analysis of the keywords in the found clusters show.

A concept that is closely related to spectral graph clustering is
Low-rank approximations of the adjacency matrix of a graph.  Tong et
al.~\cite{tpsyf-c-08} do not provide a stand-alone community detection
algorithm but a fast algorithm that returns a good low-rank
approximation of the adjacency matrix of a graph that requires only
few space.  Additionally, they propose efficient updates of these
matrix approximations that may enable many clustering methods that use
low-rank adjacency matrix approximations to also operate on evolving
graphs. 

\paragraph{Modularity Based Algorithms.}
All dynamic community detection algorithms based on explicit
modularity optimization are modifications of one of three static
agglomerative algorithms that greedily optimize modularity.

The first of these static algorithms, commonly called \algo{CNM} according to
its authors Clauset, Newman and Moore~\cite{cnm-fcsln-04}, is similar
to traditional hierarchical clustering algorithms used in data mining,
such as single linkage~\cite{s-slink-73}.  Starting from a singleton
clustering, i.e., a clustering where each cluster consists of exactly
one vertex, among all clusterings that can be reached by merging two
of the clusters, the one with the best modularity is chosen.  This is
iterated until modularity cannot be further improved by merging any of
the clusters.  Although modularity is an inherently global measure,
the improvement of the objective function after a merge operation can be easily
calculated by only considering the affected clusters.  This means that
the set of all possible merge operations can be maintained in a heap,
which leads to a total running time of $O(n^2 \log n)$.

Dinh et al.~\cite{dsttz-agami-10,dyt-t-09} evaluate a straightforward
dynamization of the \algo{CNM} algorithm that works as follows.  The graph at
the first time step is clustered with the static algorithm and the
resulting clustering is stored.  In the next time step,
we first incorporate all changes in the graph.  Then, each vertex that
is either newly inserted or incident to an edge that has been modified
is \emph{freed}, i.e., it is removed from its cluster and moved to a
newly created singleton cluster.  To arrive at the final clustering,
\algo{CNM} is used to determine if merging some of the clusters can again
improve modularity.  The authors call this framework ``Modules
Identification in Evolving Networks'' (\algo{MIEN}).

\begin{figure}[tb]
\begin{center}
  \subfloat[Example
  dendrogram]{\label{fig:CNM_dendro_a}\includegraphics[width=0.3\textwidth,page=1]{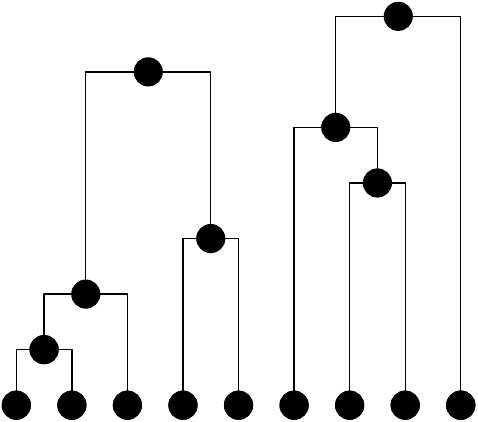}}\quad
  \subfloat[Backtracking
  in~\cite{bbp-fcddc-11}]{\includegraphics[width=0.3\textwidth,page=2]{CNM_dendro.pdf}}\quad
  \subfloat[Backtracking
  in~\cite{gmssw-dgccm-11a}]{\includegraphics[width=0.3\textwidth,page=3]{CNM_dendro.pdf}}
\end{center}
\caption{Example dendrogram and illustration of backtracking procedure
  by G\"{o}rke et al.~\cite{gmssw-dgccm-11a} and Bansal et
  al.~\cite{bbp-fcddc-11} in case an intracluster edge between the white vertices is
  deleted.}
\label{fig:CNM_dendro}
\end{figure}

Independently, G\"{o}rke et
al.~\cite{gmssw-dgccm-11a}~build upon the same idea, but in a more general setting, which results in the algorithm \algo{dGlobal}.
There are two variants of this algorithm, the first one based on freeing vertices in the neighborhood of directly affected vertices and the second one based on a \emph{backtracking} procedure.
In the first variant, the subset of freed vertices can be all vertices in the same cluster, vertices
within small hop distance or vertices found by a bounded breadth first
search starting from the set of affected vertices.
In their
experiments, considering these slightly larger subsets instead of only directly affected vertices
improves modularity and yields a good tradeoff between running time
and quality.
The second variant not only stores the clustering from the last
time step but the whole sequence of merge operations in the form of a
\emph{dendrogram}.  A dendrogram is a binary forest where leaves
correspond to vertices in the original graph and vertices on higher
levels correspond to merge operations.  Additionally, if a vertex in
the dendrogram is drawn in a level above another vertex, this encodes
that the corresponding merge has been performed later in the
algorithm.
Figure~\ref{fig:CNM_dendro_a} shows an example of a
dendrogram produced by the static \algo{CNM} algorithm whose resulting
clustering consists of two clusters.  Storing the whole dendrogram
across time steps makes backtracking strategies applicable.
To update the clustering for the next time step, the backtracking procedure first retracts a
minimum number of merges such that certain requirements are met, which depend on the type of change.
In case an intracluster edge has been inserted, the requirement is that its incident vertices are in separate clusters after the backtracking procedure.   
If an intercluster edge is inserted or an intracluster edge deleted, merges are retracted until both affected vertices are in singleton clusters.
If an intercluster edge is deleted, the dendrogram stays unchanged.  Afterwards, \algo{CNM} is used to
complete this preliminary clustering.
Bansal et al.~\cite{bbp-fcddc-11} use a similar approach.
The main difference is that instead of backtracking merges in the dendrogram, their algorithm repeats all merge operations from the last time step until an affected vertex is encountered. Again, this preliminary clustering is completed
with the static \algo{CNM} algorithm.
Figure~\ref{fig:CNM_dendro} illustrates the difference between the two
approaches.  Both studies report a speedup in running time compared to
the static algorithm, G\"{o}rke et al.~additionally show that their
approach improves smoothness significantly.  In the experiments of
Bansal et al., quality in terms of modularity is comparable to the
static algorithm, while G\"{o}rke et al.~even observe an improvement
of quality on synthetic graphs and excerpts of coauthor graphs derived
from arXiv.  G\"{o}rke et al. additionally compare the backtracking
variant of \algo{dGlobal} to the variant freeing subsets of vertices;
for the test instances, backtracking was consistently faster but
yielded worse smoothness values.

\begin{figure}[tb]
\begin{center}
  \subfloat[\algo{Louvain}]{\includegraphics[width=0.45\textwidth]{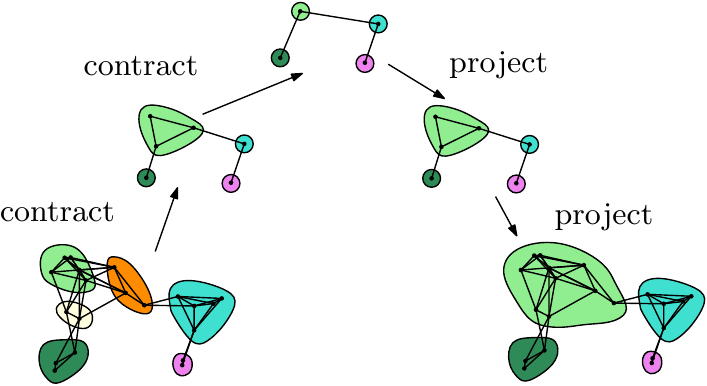}}\quad
  \subfloat[Dendrogram]{\includegraphics[width=0.4\textwidth]{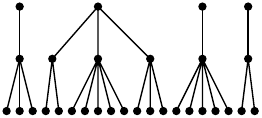}}
\end{center}
\caption{Illustration of the \algo{Louvain} method and the corresponding
  dendrogram. In the left part, the depicted edge structures show the graphs before the
  vertex moves, while the colored subsets depict the resulting clusters after
  the vertex moves on the particular level. 
}
\label{fig:Louvain}
\end{figure}

The second static algorithm that has been modified for the dynamic
scenario is a local greedy algorithm often called \algo{Louvain}
method~\cite{bgll-f-08}.  Similar to \algo{CNM}, the algorithm starts with
a singleton clustering.  Now, vertices of the graph are considered in
a random order.  If there is at least one cluster such that moving the
current vertex $v$ to it improves the overall modularity, $v$ is moved
to the cluster that yields the maximal gain in modularity.  This process is repeated
in several rounds until a local maximum is attained.  Then, clusters
are contracted to supernodes and edges between clusters summarized as
weighted edges, whereas edges within clusters are mapped to (weighted) self loops.  The
local moving procedure is then repeated on the abstracted graph taking edge
weights into account.  Contractions and vertex moves are iterated until the graphs
stays unchanged.  Then, the clustering is \emph{projected} down to the
lowest level, which represents the original graph, to get the final result.
Figure~\ref{fig:Louvain} illustrates this procedure.

Among the modifications of the \algo{Louvain} method to the dynamic scenario, the
one by Aynaud and Guillaume~\cite{ag-s-10} is the most direct.  In
their study, instead of the singleton clustering, the clustering from
the last time step is used to initialize the clustering on the lowest
level.  Using a dynamic network of webblogs, they demonstrate that
this modification improves smoothness significantly. 
In terms of modularity, the modified version follows the static version quite well and yields better quality than a reference algorithm based
 on random walks called \algo{Walktrap}~\cite{pl-cclnu-06}.
The authors further propose to use a tradeoff between modularity and
smoothness by removing a fixed percentage of randomly chosen vertices
from their cluster in each time step, in order to give the algorithm
more freedom to perform necessary changes in the clustering.  

An evolutionary version of the \algo{Louvain} method is proposed by G\"{o}rke et
al.~\cite{gmssw-dgccm-11a}, called \algo{tdLocal}.  Here, the clustering is
again reinitialized by the singleton clustering in each time step.
Inspired by Chakrabarti et al.~\cite{ckt-ec-06}, smoothness is encouraged by optimizing a linear
combination of modularity and the graph theoretic
Rand index.  It is
possible to optimize this modified objective with the \algo{Louvain}
algorithm without increasing the asymptotic running time of one round.

A dynamic version of the \algo{Louvain} method based on local updates is the algorithm QCA
proposed by Nguyen et al.~\cite{ndyt-a-11}.  Depending on the kind of
change, the following case distinction is used.
If a new vertex $v$ is inserted,  $v$ is greedily assigned to a cluster such that modularity is optimized.
In case an intercluster edge between $u$ and $v$ is inserted, the algorithm first checks if $u$ or $v$ can be moved to the other cluster such that modularity increases. 
If yes, it checks if neighbors of the moved vertex can be moved as well.
In case a vertex is deleted, its cluster is potentially split by using a method similar to clique percolation~\cite{pdfv-u-05} restricted to the affected cluster.
If an intracluster edge between $u$ and $v$ is deleted in cluster $C$,
   where $u$ and $v$ have degree at least $2$, the set of
   maximal quasi-cliques within $C$ is determined and the clustering completed similar to static \algo{CNM}.
In all other cases, the clustering from the last time step is left
unchanged.

All of
these approaches only maintain one clustering across time steps, namely, the one that could not be improved in terms of modularity.
This clustering corresponds to the highest level in the dendrogram.  In contrast
 to that, G\"{o}rke et al.~\cite{gmssw-dgccm-11a} propose to store and
possibly modify the whole dendrogram during the course of the
algorithm, which leads to the algorithm framework \algo{dLocal}.  After all
changes have been incorporated into the graph of the lowest level (the
original graph), all affected vertices, i.e., vertices that are
either new or incident to edge changes, are marked.  Additionally,
depending on some policy $P$, some vertices in their close neighbourhood
are marked as well.  The set of policies evaluated
in this study correspond to the aforementioned subset strategies
evaluated by G\"{o}rke et al.~\cite{gmssw-dgccm-11a} for \algo{dGlobal}.
This means, $P$ can correspond to freeing vertices within a small hop
distance from affected vertices, vertices in the same cluster and
vertices found by a bounded breadth first search.  Then, vertices on
the lowest level are greedily moved until modularity cannot be further
improved.  Now, affected vertices in the second level of the
dendrogram are marked, i.e., subclusters affected by lower level
changes.  Depending on $P$, again, some vertices in their neighborhood
are marked and possibly moved.  This is repeated on all levels.
The current clustering can be found at all time steps by considering
the projection of the highest to the lowest level.  Keeping the whole
dendrogram in memory instead of only its highest level yields the
additional possibility to merge and split clusters on intermediate
levels and therefore increases the search space used for local moving,
which leads to possibly better quality.
Compared to the algorithm \algo{dGlobal}, the experiments of G\"{o}rke et al.~do not give a conclusive results; which of these algorithms performs better depends on the structure of the graph and its communities.

An approach that is very similar to \algo{dLocal} is used in the algorithm $\text{A}^3$CS proposed by Nguyen et al.~\cite{dnt-aaaac-13}.
The main difference is that the lowest level of the dendrogram is not computed and maintained by a local moving procedure but by an algorithm similar to the static Low-degree Following Algorithm proposed by Dinh and Thai~\cite{dt-cdsfn-13}.
This algorithm has the nice property to yield approximation guarantees for graphs with a perfect powerlaw degree distribution with a powerlaw exponent larger than $2$.
This property is inherited by the dynamic algorithm $\text{A}^3$CS.
However, the clusters produced by this algorithm are quite small, which is why it makes sense to additionally consider local moving (which includes merging of clusters) on higher levels to further improve its practical performance.
Dinh and Thai compare $\text{A}^3$CS to QCA and MIEN, with the result that it dominates both with respect to running time and quality.

The third static algorithm for modularity based clustering, which
lends itself especially well to parallelization, is based on the
contraction of matchings~\cite{rn-m-11,sc-emomg-08}.  To effectively
optimize modularity, edges are weighted based on the gain in
modularity corresponding to the merge of the two adjacent vertices.
Using these edge weights, a weighted matching is computed in a greedy
fashion, possibly in parallel~\cite{rmeb-pcdmg-12}.  Riedy and
Bader~\cite{rb-mcmms-13} propose a dynamic variant of this algorithm
especially for the case of larger batch sizes, i.e., many changes
between consecutive time steps.  Across the time steps, the current
clustering together with its community graph is stored.  After
incorporating the edge changes in the community graph, all
(elementary) vertices incident to newly inserted intercluster or
deleted intracluster edges are extracted from their community.  Then,
the matching based agglomeration is restarted from the modified
community graph.  As the community graph is usually much smaller than
the original graph, this potentially saves running time.

\paragraph{Label Propagation/Diffusion.}
An algorithm that is not based on modularity, but conceptually closely
related to the \algo{Louvain} method is \algo{Label
  propagation}~\cite{rak-n-07}.  \algo{Label propagation} can be seen
as an algorithm greedily maximizing the number of edges within
clusters by moving vertices, very similar to the local moving
procedure in the \algo{Louvain} method.  Obviously, the global optimum
with respect to the number of edges within clusters is trivial, as it
corresponds to assigning all vertices to one cluster.  Hence, using
local moving in a multilevel scheme, as in the case of modularity
maximization, does not make much sense.  Instead, one is interested in
the local maximum obtained after local moving on the original graph,
which corresponds to the lowest level in the dendrogram.  \algo{Label
  propagation} is very fast and lends itself well to
parallelization~\cite{sm-ehpcd-13}.  Xie and Szymanski propose a
modified version of this algorithm called
\algo{LabelRank}~\cite{xs-lrasl-13}.
In contrast to the original \algo{Label propagation} algorithm, each
vertex does not maintain one single cluster id or \emph{label}, but
instead a vector of containment probabilities for each cluster
currently existing in the graph.  Label propagation phases alternate
with inflation and cutoff steps to decrease the computational
complexity and to make the differences in the particular vectors more
pronounced.  To prevent the algorithm from converging too fast to the
(potentially uninteresting) static distribution, only labels of
vertices are updated that are sufficiently different from their
neighbors.  The algorithm outputs a set of labels for each vertex that
each has an associated probability, which would lead to overlapping
clusters.  This is why, although overlaps are resolved in a
preprocessing step by only considering the strongest label for each
vertex, we list the algorithm both among overlapping and non
overlapping approaches in Table~\ref{tab:properties_dyn}.  Both
\algo{Label propagation} and \algo{LabelRank} have been modified to
the dynamic scenario~\cite{pcw-a-09,xcs-lrtic-13}, roughly based on
the idea to only update labels/label vectors of vertices affected by
changes in the graph.  The dynamic version of \algo{LabelRank} is
called \algo{LabelRankT}.

A concept that is very similar to \algo{LabelRank} and has been developed in
the context of graph partitioning is
\emph{diffusion}~\cite{gs-igpis-06,mms-a-09,mms-gpdd-09}.  Similar to
the above algorithm, each vertex maintains a vector of size $k$
indicating to which extent it is connected to the vertices of each of
the $k$ clusters.  The entries of these vectors are called
\emph{loads}; loads are distributed through the network along the
edges in rounds, which explains the origin of the name diffusion.
Based on this concept, Gehweiler and Meyerhenke~\cite{gm-addhc-10}
propose a distributed graph clustering algorithm called \algo{DiDiC}, which
is motivated by the task to cluster nodes of a peer-to-peer based
virtual distributed supercomputer.  The weight of edges between nodes
in this network corresponds to the bandwidth between the associated
peers. 
The idea is to find clusters of highly connected peers that can
be used to solve a common task in parallel.  In contrast to
\algo{LabelRankT}, they use a second diffusion system drawing the loads associated with
cluster $i$ back to the vertices in cluster $i$, which accelerates the
formation of large, connected clusters.  In the first time step, the
process starts with a random clustering and distributes the load of
each cluster to the vertices it contains.  After the diffusion process
has been run for a certain number of rounds, clusters are reassigned
such that each vertex moves to the cluster from which it obtained the
highest load value, leading to a complete clustering.  The algorithm
is made dynamic by initializing the clusters and load vectors with the
values obtained in the previous time step, instead of random
initialization.

\paragraph{Generative Models.}
Another popular approach to clustering problems is the use of
generative models that assume the graph is randomly generated on the
basis of a hidden ground truth clustering.  The goal is now to
approximately recover the hidden or latent clustering by looking for
clusterings that are \emph{likely} given the observed outcome of this
random process, which corresponds to the given graph.
Given conditional probabilities that estimate this likelihood and a
prior distribution over the set of all possible clusterings, the
posterior probability of a given clustering can be obtained via Bayes'
theorem.
For conditional probabilities, a common choice are \emph{stochastic
  block models}~\cite{awf-bsb-92,sn-epsb-97,ww-sbdg-87}, which
generally assume that the probability of a link between two vertices
is determined by their cluster membership.  If the number of clusters
is not provided as an input parameter, a popular choice as a prior
distribution for the cluster assignments is the distribution induced
by the Chinese restaurant process~\cite{a-e-85} and its derivations.
The goal of maximum a posteriori (MAP) estimations is now to find
solutions with high posterior
probability.

Among the few approaches based on these concepts that explicitly
address the issue of dynamic graphs are
\algo{FacetNet}~\cite{lczst-a-09} and the algorithm by Yang et
al.~\cite{yczj-db-11}.  The goal of both approaches is to implicitly
enforce smoothness by choosing the prior distribution such that large
changes in the clustering between adjacent time steps are assumed to
be unlikely.  In contrast to traditional stochastic block models,
\algo{FacetNet} builds upon the model proposed by Yu et
al.~\cite{yyt-scg-06} that assumes ``soft community membership'',
i.e.~vertices belong to different clusters to more or less extent.
This results in an overlapping clustering.  However, these clusters
can easily be converted to a complete clustering in a postprocessing
step by assigning each vertex to the cluster it participates in to the
largest extent.  For this reason and the fact that this is often done
when comparing complete clusterings to the clusterings produced by
\algo{FacetNet}, we list the algorithm both under overlapping and
non overlapping clusterings in
Table~\ref{tab:properties_dyn}.  In the generative model, the
probability of a certain cluster assignment at time step $t$ depends
on the cluster assignment at step $t-1$.  Depending on a parameter
$\nu$, the transitions will be more or less smooth.  It can be shown
that under certain assumptions, the MAP estimation of
this model is equivalent to the framework of
Chakrabarti~\cite{ckt-ec-06}.  In this context, the KL-divergence
between the observed weight matrix and an approximation of it based on
cluster assignments is used as the snapshot cost and the KL-divergence
between the clustering at time step $t$ and at time step $t-1$ as
history cost.  For the inference step, an expectation maximization
algorithm is used that is guaranteed to converge towards a locally
optimal solution of the corresponding MAP problem.  In the
\algo{FacetNet} framework, the number of clusters can change over
time.  To determine the best number of clusters for each time step, an
extension of modularity to soft community memberships is proposed.  In
the experimental part, synthetic and real world networks are used to
evaluate the performance of \algo{FacetNet} and to compare it to its
static counterpart as well as a static and evolutionary
(\algo{EvolSpec}) version of spectral
clustering~\cite{cszht-e-07,sm-ncis-00}.  With respect to quality, the
\algo{FacetNet} approach compares favorably.

In the algorithm of Yang et al., the number of clusters is given as
input.  Given the hidden clustering at a certain time step, the
conditional probability for a link between two vertices is determined
by the linking probabilities associated with their respective
clusters.  These linking probabilities are in turn random variables
such that their prior distribution causes higher linking probabilities
for intracluster edges.  The whole generative model corresponds to a
Bayesian net where the latent variables associated with a certain time
step depend on the clustering from the last time step, a matrix $A$
specifying the probability that a vertex moves from a certain cluster
to another in the current time step, and the prior distribution for
the linking probabilities between clusters.  Again, the prior
distribution for $A$ biases the moving probabilities in such a way
that the probability for each vertex to move to another community $k$
is smaller than the probability to stay in its own cluster, which
implicitly biases the model towards smoothness.  The model can be
generalized to weighted graphs in a straightforward way.  For the
inference step, the authors evaluate both the online and the offline
scenario.  In the online scenario, the variables are sampled from time
step to time step using the observations seen so far.  In the offline
scenario, all variables are sampled together by taking both past and
future observations into account.  In both cases, a Gibbs sampler is
used to infer the latent variables.  In the offline scenario,
additionally, an expectation maximization algorithm is proposed.
These two variants are then compared against each other, against
static statistical blockmodels, and the dynamic algorithms
\algo{EvolSpec} and \algo{FacetNet} and their static counterparts.
Experiments on synthetic and real world networks suggest that the
approach based on Gibbs sampling in combination with the new
generative model yields the best quality.  It might be worth to
mention that the hyperparameters that influence the priors are tuned
by considering the modularity of the resulting clusterings.

Another generative approach has been used by Sun et
al.~\cite{sthgz-c-10} for their algorithm \algo{Evo-NetClus}.  Similar
to \algo{FacetNet}, \algo{Evo-NetClus} builds upon soft community
membership.  In contrast to the models mentioned above, the networks
considered are assumed to contain vertices of multiple types, where
one of the types is defined as a special ``center type''.  Each edge
is incident to exactly one center type vertex; the authors call this
star network schema.  As an illustrative example that is also used in
the experiments, the authors use publication networks as for example
DBLP\footnote{\url{http://www.informatik.uni-trier.de/~ley/db/}}.
Here, vertices of the center type correspond to papers and the other
types to authors, conferences and associated terms.  Again, the
probability of the clustering of time step $t$ is influenced by the
clustering at time step $t-1$, favoring smoothness.  The model
incorporates changing cluster numbers in each time steps that are not
assumed to be input parameters.  For the inference step, an online
Gibbs sampler is proposed.  With respect to quality, the authors
compare their model to degenerated models that do not take historical
data or only a subset of types into account.

\paragraph{Further Approaches.}
In this category we list three approaches that do not fit into any of
the previous categories.  The first approach considers a bicriterial
optimization problem while the former approaches focus on a single
criterion, the second approach is a general framework that allows to
incorporate temporal smoothness into basically every static clustering
algorithm, and the third approach claims that considering the input
graph as a homogeneous structure where in each region the same criteria
for good clusters hold is not appropriate.

The approach of Kim et al.~\cite{kmm-m-10} is based on optimizing two
different measures simultaneously in a bicriterial approach.
Typically, the measures in bicriterial approaches are competing in the
sense that one measure tends towards the 1-clustering and the other
towards the singleton clustering.  The goal is to approximate the
\emph{pareto front}, i.e., to find a set of clusterings that are not
dominated by other clusterings with respect to both criteria.  Kim et
al.\ use as criteria (or \emph{fitness functions}) a
global version of MinMaxCut~\cite{dhzgs-amcag-01}, which tends to the
1-clustering, and a global version of the silhouette
index~\cite{r-sa-87}, which tends to the singleton clustering.  They
approximate the pareto front by an evolutionary metaheuristic in a
dynamic scenario using a \emph{locus-based representation} of
clusterings~\cite{ps-agacp-98}, which is a vector of length $n$
storing for each vertex exactly one outgoing edge.  The represented
clustering then corresponds to the connected components of the induced
graph.  The locus-based representation has the advantage that
different clusterings can be combined (crossed) in a meaningful way by
performing \emph{uniform crossover} on the corresponding vectors,
which means that each entry in the resulting vector is randomly taken
from one of the parent vectors.  The dynamization is trivially done by
initializing the population of the current time step by the result
from the last time step.  Different evolutionary metaheuristics are
compared with respect to both criteria on a dynamic graph representing
YouTube videos.

While the former approach uses evolutionary metaheuristics, which have
nothing to do with evolutionary clustering according to Chakrabarti et
al.~\cite{ckt-ec-06}, the next approach is again an evolutionary
clustering approach. In contrast to other evolutionary clustering
approaches, which most often incorporate temporal smoothness into a
particular clustering algorithm,
the framework introduced by Xu et al.~\cite{xki-tcdsn-11} can be
applied with any static clustering method.
In their publication the authors use the normalized cut spectral
clustering approach by Yu and Shi~\cite{ys-m-03}.
Although the idea of Xu et al.\ is inspired by Chakrabarti et al., the
main difference is that they do not incorporate temporal smoothness by
optimizing a linear combination of snapshot quality and history
quality, but adapt the input data for the chosen clustering algorithm
based on the community structure found in the previous snapshot.  This
adaption is done as follows.  The adjacency matrices of the snapshots
are considered as realizations of a non stationary random process which
allows to define an expected adjacency matrix for the current
snapshot.  Based on this expected matrix a \emph{smoothed adjacency
  matrix} can be approximated that also takes into account the
previous time step.  The smoothed adjacency matrix is a convex
combination of the smoothed adjacency matrix of the previous time step
and the actual adjacency matrix of the current time step.  The
parameter that balances the two terms of the convex combination is
estimated such that it minimizes a mean squared error criterion.  The
chosen clustering algorithm is then applied to the estimated smoothed
adjacency matrix, thus incorporating temporal smoothness to stabilize
the variation of the found clusters over time.

All the above clustering approaches use the same objective for the
whole graph to get good clusterings.  In this way these approaches
consider the input graph as homogeneous structure, regardless whether
parts of the graph are sparser than others, and thus, possibly require
another notion of density for reasonable communities than denser
parts.  Wang et al.~\cite{lwy-dcdwg-13} follow Aggarwal et
al.~\cite{axy-tcdlh-11} who claim that considering networks as homogeneous
structures is not an appropriate attempt.  This is why Wang et al.\
introduce patterns describing homogeneous regions that are consolidated
in a second step to generate non overlapping clusters.  In contrast to
density, which depends on the number or the weight of edges within a
subgraph or cluster, homogeneity means that all vertices in a pattern
have similarly weighted neighbors. In order to efficiently compute the
patterns in a dynamic scenario, the authors maintain, by incremental
updates, a top-$k$ neighbor list and a top-$k$ candidate list as
auxiliary structures.  These updates are able to deal with atomic
changes as well as with several changes (of vertices and edge weights)
in one time step.
In comparison with
\algo{FacetNet}~\cite{lczst-a-09} and the evolutionary clustering
method by Kim and Han~\cite{kh-a-09}, experiments on the DBLP, the ACM
and the IBM data set prove a better processing rate (number of
vertices processed per second) and a better accuracy of the found
clusterings in locally heterogeneous graphs.


\paragraph{Summary.}
To provides a summary of all algorithms described in this
section, Table~\ref{tab:properties_dyn} lists some of their basic
properties.  These include whether the authors aim at running time or
smoothness, or both, and if the resulting clusterings  are overlapping or
not. If applicable, we further give a reference to an existing static
algorithm the approach is based upon.  Among the algorithms we
considered, the majority focuses on the task of finding non
overlapping clusters.  Interesting is that the number of algorithms
aiming at low running time is almost the same as the number of algorithms
aiming at smoothness; only very few algorithms take both into account.
Note that an entry in the table indicating that an algorithm does not
aim at low running time or smoothness does not 
indicate that the
algorithm is slow or produces unsmooth clusterings; it just signifies
that this aspect was neither considered in the conception of the
algorithm nor evaluated in the experiments.  In general, it is
expected that smoothness and running time go together quite well, as the
use of local updates often improves both of these aspects.

\renewcommand{\arraystretch}{1.3} \setlength{\tabcolsep}{0.9mm}
\begin{table}[tbp]
  \caption{Systematic overview on main features of the clustering algorithms and
    community detection approaches presented in
    Section~\ref{sec:IncrClus}.}
\label{tab:properties_dyn}
 \begin{center}
\begin{footnotesize}
  \begin{tabular}{l|c|c|c|c|l}
    \hline
    & \multicolumn{2}{c|}{aims at} & \multicolumn{2}{c|}{overlapping}
    & \multicolumn{1}{c}{based on}\\
    Reference & run.~time & smoothn. & yes & no & 
    \multicolumn{1}{c}{existing static approach}\\
    \hline \hline
    \multirow{2}{*}{\vspace{1ex} \hspace{-0.5ex}\parbox[t][2em][t]{35.9mm}{
        Sun et al.~\cite{sypf-gspfm-07} 
        \hspace*{0ex}\scriptsize (\algo{GraphScope})}
    }
    & $\no$ & $\yes$ & $\no$ & $\yes$ & $\no$\\ 
    \hline
    \multirow{2}{*}{\vspace{1ex} \hspace{-0.5ex}\parbox[t][2em][t]{35mm}{        
        Angel at al.~\cite{asks-d-12} 
        \hspace*{0ex}\scriptsize (\algo{DynDens})}
    }
    & $\yes$ & $\no$ & $\yes$ & $\no$ & $\no$ \\
    Agarwal et al.~\cite{arb-r-12} 
    & $\yes$ & $\no$ & $\yes$ & $\no$ & $\no$ \\ 
    \hline
    Takaffoli et al.~\cite{trz-ilcid-13}
    & $\no$ & $\yes$ & ($\checkmark$) & $\yes$ & Chen et al.~\cite{czg-dclni-09}\\
    Kim and Han~\cite{kh-a-09}
    & $\no$ & $\yes$ & ($\checkmark$) & $\yes$ & 
    \multirow{2}{*}{\vspace{1ex} \hspace{-0.5ex}\parbox[t][2em][t]{40mm}{  
        Xu et al.~\cite{xyfa-scana-07}
        \hspace*{0ex}\scriptsize (\algo{Scan})}
    }\\
    \multirow{2}{*}{\vspace{1ex} \hspace{-0.5ex}\parbox[t][2em][t]{35mm}{
        Falkowski et al.~\cite{fbs-dengr-07,f-cacad-09} 
        \newline {\hspace*{0ex}\scriptsize (\algo{DenGraph})}} } 
    & $\yes$ & $\no$ & $\yes$ & $\yes$ & 
    \multirow{2}{*}{\vspace{1ex} \hspace{-0.5ex}\parbox[t][2em][t]{40mm}{  
        Xu et al.~\cite{xyfa-scana-07}
        \hspace*{0ex}\scriptsize (\algo{Scan})}
    }\\
    &&&&&\\ 
    Cazabet et al.~\cite{cah-d-10} 
    & $\no$ & $\yes$ & $\yes$ & $\no$ & $\no$ \\ 
    \hline
    Duan et al.~\cite{dyll-ik-12}
    & $\yes$ & $\no$ & $\yes$ & $\no$ & 
    \multirow{2}{*}{\vspace{1ex} \hspace{-0.5ex}\parbox[t][2em][t]{40mm}{  
        Der\'enyi et al.~\cite{dpv-cprn-05}
        \hspace*{0ex}\scriptsize (\algo{PCM})}
    }\\
    G\"orke et al.~\cite{ghw-dmctc-09,ghw-dcc-11}
    & $\yes$ & $\yes$  & $\no$ & $\yes$  & Flake et al.~\cite{ftt-gcmct-04}\\ 
    \hline
    \multirow{2}{*}{\vspace{1ex} \hspace{-0.5ex}\parbox[t][2em][t]{35mm}{
       Chi et al.~\cite{cszht-e-07}
        \hspace*{0ex}\scriptsize (\algo{EvolSpec})}
    }
    & $\no$ & $\yes$ & $\no$ & $\yes$ &  Shi and Malik~\cite{sm-ncis-00}\\ 
    Ning et al.~\cite{nxcgh-i-10}
    & $\yes$ & $\no$ & $\no$ & $\yes$ & Shi and Malik~\cite{sm-ncis-00}\\ 
    \hline
    \multirow{2}{*}{\vspace{1ex} \hspace{-0.5ex}\parbox[t][2em][t]{35mm}{
        Dinh et al.~\cite{dsttz-agami-10,dyt-t-09}
          \hspace*{0ex}\scriptsize (\algo{MIEN})}
    }
    & $\yes$ & $\no$ & $\no$ & $\yes$ & 
    \multirow{2}{*}{\vspace{1ex} \hspace{-0.5ex}\parbox[t][2em][t]{40mm}{  
        Newman and Moore~\cite{cnm-fcsln-04}
        \hspace*{0ex}\scriptsize (\algo{CNM})}
    }\\
    \multirow{2}{*}{\vspace{1ex} \hspace{-0.5ex}\parbox[t][2em][t]{35mm}{
        G\"{o}rke et al.~\cite{gmssw-dgccm-11a}
        \hspace*{0ex}\scriptsize (\algo{dGlobal})}
    }
    & $\yes$ & $\yes$ & $\no$ & $\yes$ & 
    \multirow{2}{*}{ \vspace{1ex}\hspace{-0.5ex}\parbox[t][2em][t]{40mm}{  
        Newman and Moore~\cite{cnm-fcsln-04}
        \hspace*{0ex}\scriptsize (\algo{CNM})}
    }\\
    Bansal et al.~\cite{bbp-fcddc-11}
    & $\yes$ & $\no$ & $\no$ & $\yes$ & 
    \multirow{2}{*}{\vspace{1ex} \hspace{-0.5ex}\parbox[t][2em][t]{40mm}{  
        Newman and Moore~\cite{cnm-fcsln-04}
        \hspace*{0ex}\scriptsize (CNM)}
    }\\
    Aynaud and Guillaume~\cite{ag-s-10}
    & $\no$ & $\yes$ & $\no$ & $\yes$ & 
    \multirow{2}{*}{\vspace{1ex} \hspace{-0.5ex}\parbox[t][2em][t]{40mm}{  
        Blondel et al.~\cite{bgll-f-08}
        \hspace*{0ex}\scriptsize (\algo{Louvain})}
    }\\
    \multirow{2}{*}{\vspace{1ex} \hspace{-0.5ex}\parbox[t][2em][t]{35mm}{        
        G\"{o}rke et al.~\cite{gmssw-dgccm-11a}
          \hspace*{0ex}\scriptsize (\algo{tdLocal})}
    }
    & $\no$ & $\yes$ & $\no$ & $\yes$ & 
    \multirow{2}{*}{\vspace{1ex} \hspace{-0.5ex}\parbox[t][2em][t]{40mm}{  
        Blondel et al.~\cite{bgll-f-08}
        \hspace*{0ex}\scriptsize (\algo{Louvain})}
    }\\
    \multirow{2}{*}{\vspace{1ex} \hspace{-0.5ex}\parbox[t][2em][t]{35mm}{    
        Nguyen et al.~\cite{ndyt-a-11}
        \hspace*{0ex}\scriptsize (\algo{QCA})}
    }
    & $\yes$ & $\no$ & $\no$ & $\yes$ & 
    \multirow{2}{*}{\vspace{1ex} \hspace{-0.5ex}\parbox[t][2em][t]{40mm}{  
        Blondel et al.~\cite{bgll-f-08}
        \hspace*{0ex}\scriptsize (\algo{Louvain})}
    }\\
    \multirow{2}{*}{\vspace{1ex} \hspace{-0.5ex}\parbox[t][2em][t]{35mm}{        
        G\"{o}rke et al.~\cite{gmssw-dgccm-11a}
          \hspace*{0ex}\scriptsize (\algo{dLocal})}
    }
    & $\yes$ & $\yes$ & $\no$ & $\yes$ & 
    \multirow{2}{*}{\vspace{1ex} \hspace{-0.5ex}\parbox[t][2em][t]{40mm}{  
        Blondel et al.~\cite{bgll-f-08}
        \hspace*{0ex}\scriptsize (\algo{Louvain})}
    }\\
    \multirow{2}{*}{\vspace{1ex} \hspace{-0.5ex}\parbox[t][2em][t]{35mm}{       
        Nguyen et al.~\cite{dnt-aaaac-13}
        \hspace*{0ex}\scriptsize (\algo{$\text{A}^3$CS})}
    }
    & $\yes$ & $\no$ & $\no$ & $\yes$ & 
    \multirow{2}{*}{\vspace{1ex} \hspace{-0.5ex}\parbox[t][2em][t]{40mm}{  
        Dinh and Thai~\cite{dt-cdsfn-13} and
        \newline {Blondel et al.~\cite{bgll-f-08} \scriptsize (\algo{Louvain})}}
    }\\
    &&&&&\\
    Riedy and Bader~\cite{rb-mcmms-13}
    & $\yes$ & $\no$ & $\no$ & $\yes$ & Riedy et al.~\cite{rmeb-pcdmg-12}\\
        \hline
    Pang et al.~\cite{pcw-a-09}
    & $\yes$ & $\no$ & $\no$ & $\yes$ & 
    \multirow{2}{*}{\vspace{1ex} \hspace{-0.5ex}\parbox[t][2em][t]{40mm}{  
        Raghavan et al.~\cite{rak-n-07}
        \newline {\vspace{1ex}\hspace*{0ex}\scriptsize (\algo{Label propagation})}} }\\
    &&&&&\\
    \multirow{2}{*}{\vspace{1ex} \hspace{-0.5ex}\parbox[t][2em][t]{35mm}{           
        Xie et al.~\cite{xcs-lrtic-13}
        \newline {\hspace*{0ex}\scriptsize (\algo{LabelRankT})}}}
    & $\yes$ & $\no$ & $\yes$ & $\yes$ & 
    \multirow{2}{*}{\vspace{1ex} \hspace{-0.5ex}\parbox[t][2em][t]{40mm}{  
        Xie et al.~\cite{xs-lrasl-13}
        \hspace*{0ex}\scriptsize (\algo{LabelRank})}
    }\\
    &&&&&\\
    \multirow{2}{*}{\vspace{1ex} \hspace{-0.5ex}\parbox[t][2em][t]{35mm}{      
        Gehweiler et al.~\cite{gm-addhc-10}
        \hspace*{0ex}\scriptsize (\algo{DiDiC})}
    }
    & $\yes$ & $\no$ & $\no$ & $\yes$ & Meyerhenke et al.~\cite{mms-gpdd-09}\\
    \hline
    \multirow{2}{*}{\vspace{1ex} \hspace{-0.5ex}\parbox[t][2em][t]{35mm}{      
        Lin et al.~\cite{lczst-a-09}
        \hspace*{0ex}\scriptsize (\algo{FacetNet})}
    }
    & $\no$ & $\yes$ & $\yes$ & ($\checkmark$) & Yu et al.~\cite{yyt-scg-06}\\
    Yang et al.~\cite{yczj-db-11}
    & $\no$ & $\yes$ & $\no$ & $\yes$ & $\no$ \\
    \multirow{2}{*}{\vspace{1ex} \hspace{-0.5ex}\parbox[t][2em][t]{35mm}{      
        Sun et al.~\cite{sthgz-c-10}
        \newline {
        \hspace*{0ex}\scriptsize (\algo{Evo-NetClus})} } 
    }
    & $\no$ & $\yes$ & $\yes$ & ($\checkmark$) &     
    \multirow{2}{*}{\vspace{1ex} \hspace{-0.5ex}\parbox[t][2em][t]{35mm}{      
        Sun et al.~\cite{syh-rchin-09}
        \hspace*{0ex}\scriptsize (\algo{NetClus})}
    }\\
    &&&&&\\
    \hline
    Kim et al.~\cite{kmm-m-10}
    & $\no$ & $\yes$ & $\no$ & $\yes$ & $\no$ \\
    Wang et al.~\cite{lwy-dcdwg-13}
    & $\yes$ & $\no$ & $\no$ & $\yes$ & $\no$\\
    Xu et al.~\cite{xki-tcdsn-11}
    & $\no$ & $\yes$ & $\no$ & $\yes$ & $\no$\\
    \hline
  \end{tabular}
\end{footnotesize}
\end{center}

\end{table}

\section{Data Sets}\label{sec:benchAndGen}
\label{sec:dataSets}
In this section, we aim to give an overview on what kind of data sets have been used in current publications regarding the clustering of evolving graphs.
In the first part, we concentrate on real world instances, i.e., instances that correspond to data collected from observed relationships between objects or persons.
In the second part, we briefly talk about models and generators for evolving networks, with a special focus on synthetic data incorporating a hidden ground truth clustering.

\subsection{Real World instances}
\label{sec:benchmarks}
Most networks described in this category are based on human interaction and can therefore be classified as \emph{social networks} in the wider sense.
We tried to assign them to more fine grained subcategories depending on their structure and interpretation.
\paragraph{Email networks.}
One of the few publicly available networks corresponding to social
interaction and containing both time information and additional
metadata is the Enron email dataset\footnote{available at
  \url{http://www.cs.cmu.edu/~enron/}}.  It represents the email
exchange of employees of the Enron Corporation and was made public
during the legal investigation concerning the Enron corporation.
According to the information given on the above mentioned website, the
dataset contains about $600000$ emails belonging to $158$ users.  Note
that the total number of distinct email addresses in the data is much
larger, as also emails to and from non-Enron email addresses are
recorded.  In most network representations of the dataset, employees
are modeled as vertices and two vertices are connected by an edge if
and only if the dataset contains an email between the two
corresponding employees.  Since the data also distinguishes between
sender and recipient, edges are sometimes directed.  Furthermore, the
emails of a certain time period are often aggregated in one snapshot.
This may result in multiple edges or in weighted edges representing
the frequency.  Hence, depending on the concrete modeling, different
authors refer to quite different dynamic graphs as ``the Enron
network'', which makes comparisons between the experimental findings
rather difficult.  This is also the case for static data sets;
however, due to even more degrees of freedom, for example the
frequency of time steps or the question whether relations age and
disappear over time, this is even more immanent in the case of dynamic
data.  Takaffoli et al.~\cite{trz-ilcid-13} choose monthly snapshots
over one year, which considerably decreases the number of vertices and
edges in the network.  As a kind of ground truth clustering, they
identify ``persisting topics'' based on keyword extraction.  Duan et
al.~\cite{dyll-ik-12} and Dinh et al.~\cite{dyt-t-09,ndyt-a-11}
consider emails on a weekly basis and do not consider any metadata.
The Enron dataset has also been used in the evaluation of
\algo{GraphScope}~\cite{sypf-gspfm-07}; however, there the data is
considered as a directed and bipartite graph, where one part
corresponds to senders and the other to receivers of emails.
G\"{o}rke et al.~\cite{gmssw-dgccm-11a} did not consider Enron data,
but an anonymized network of e-mail contacts at the department of
computer science at KIT.  It is comprised of about $700000$ events
collected over a period of about $2.5$ years\footnote{For further
  details and for downloading the whole dataset, please visit
  \url{http://i11www.iti.uni-karlsruhe.de/en/projects/spp1307/emaildata}}.
As metadata, it includes an id for each email address specifying the
corresponding chair, which can be considered as ground truth clusters.

\paragraph{Cellphone data.}
Very similar to email networks are data about phone calls.  Palla et
al.~\cite{pbv-q-07} cluster a network of phone calls between the
customers of a mobile phone company containing data of over $4$
million users.  They consider edges to be weighted; a phone call
contributes to the weight between the participating customers for some
time period around the actual time of the call.
As metadata to evaluate their community finding approach, they
consider zip code and age of customers.  Similar data is considered by
Green et al.~\cite{gdc-tecds-10}, however, they do not consider edge
weights.  The \emph{Reality MiningDataset}~\cite{ep-r-06} is provided
by the MIT Human Dynamics
Lab\footnote{\url{http://realitycommons.media.mit.edu/realitymining.html}}
and was collected during a social science experiment in 2004.
It includes information about call logs, Bluetooth devices in
proximity, cell tower IDs, application usage, and phone status of $94$
subjects over the course of an academic year.  In the context of
dynamic graph clustering, it is possible to extract test data in
various ways.  Xu et al.~\cite{xki-tcdsn-11} construct a dynamic graph
where the edge weight between two participants in a time step
corresponds to the number of intervals in which they were in close
physical proximity.  As ground truth clustering, they use the affiliations of the
participants.  Sun et al.~\cite{sypf-gspfm-07} additionally consider
the cellphone activity to construct a second dynamic graph.

\paragraph{Online social networks and blogs.}
Another prime example of social networks are online social networks
like Facebook or Flickr.  In the context of clustering algorithms,
they are particularly interesting due to their size and the fact that
friendship links are explicit and not implicitly assumed with the
help of other metadata.  Viswanath et al.~\cite{vmcg-oeuif-09} crawled
the regional network of Facebook in New Orleans.  Only data from
public profiles is collected, giving information about approximately
$63000$ users and $1.5$ Mio. friendship links, together with their
evolution.  Nguyen et al.~\cite{ndyt-a-11} and Dinh et
al.~\cite{dnt-aaaac-13} use these data to evaluate their clustering
algorithms.  Kumar et al.~\cite{knt-s-06} analyze data from
Flickr\footnote{\url{http://www.flickr.com/}} and Yahoo! $360^\circ$.
Whereas Yahoo! $360^\circ$ was a typical social network that does not
exist anymore, Flickr has a focus on the sharing of photos, although
friendship links exist as well.  Both datasets
 are used by the
authors in anonymized form and are not publicly available.
Closely related to online social networks are networks derived from blogging platforms; here, the edges correspond to entry-to-entry
links between different blogs~\cite{ag-s-10,lczst-a-09,nxcgh-i-10,yczj-db-11}.  Angel et
al.~\cite{asks-d-12} use sampled data obtained via Twitter's
restricted access to its data
stream\footnote{\url{https://dev.twitter.com/docs/streaming-apis\#sampling}}.
LiveJournal\footnote{\url{http://www.livejournal.com/}} is somewhere
in between a network of blogs and an online social network.
Interesting is that users can explicitly create friendship links as
well as join groups.  In contrast to the usual way dynamic networks
are build from blogs, edges do not necessarily correspond to links but
can also depend on friendship links.  Backstrom et
al.~\cite{bhkl-g-06} study the evolution of communities in LiveJournal
using friendship links as edges and group membership as (overlapping)
ground truth clustering.  

\paragraph{Publication databases.}
Publication databases can be used to extract dynamic graphs in several
ways.  Probably the most common approach is to consider \emph{coauthor
  graphs}, in which vertices correspond to authors, and two authors
are linked if they coauthored at least one publication.  Depending on
the model, edges are weighted in different ways depending on the
number of shared publications and the number of authors on each
publication.  An orthogonal view on the data yields \emph{copaper
  networks} where vertices correspond to papers and links exist if
papers have at least one author in common.  Both of these network
types are simplifications of bipartite \emph{author-paper} networks
that relate authors to their articles.  Another possibility is to not
take authorship into account but insert (possibly directed) links
between articles if one article cites the other, leading to
\emph{citation networks}.  It is commonly believed that clusters in
all of these networks correspond to different research topics or
fields.  Due to the fact that publication data is typically not
subject to any privacy concerns and their size is reasonably large,
they are often used in the evaluation of graph clustering
algorithms~\cite{apu-a-09,ag-s-10,bhkl-g-06,bbp-fcddc-11,dnt-aaaac-13,dyt-t-09,dyll-ik-12,gmssw-dgccm-11a,lczst-a-09,ndyt-a-11,sthgz-c-10,tfsz-t-11,xcs-lrtic-13}.
Another advantage is that information about conferences and journals
the articles appeared in can be used as metadata to evaluate the
resulting clusterings.  The temporal aspect in the data stems from the
fact that each publication has an associated publication year.  The
two most often considered databases in the context of clustering are
DBLP and arXiv.  DBLP collects information about publications in the
field of computer science; information about how this data can be
downloaded as an xml file can be found on the corresponding
homepage\footnote{\url{http://www.informatik.uni-trier.de/~ley/db/}}.
The arXiv e-print archive\footnote{\url{http://arxiv.org/}} is a
repository that stores electronic e-prints, organized in several
categories alongside time stamped metadata.  To evaluate their dynamic
graph clustering framework, G\"{o}rke et al.~\cite{gmssw-dgccm-11a}
used a dataset obtained from this repository, which can be found,
together with the source code of the crawler used to extract this
data, on the corresponding project
page\footnote{\url{http://i11www.iti.uni-karlsruhe.de/en/projects/spp1307/dyneval}}.
The KDD cup 2003 also provides further arXiv datasets on its project
page\footnote{\url{http://www.cs.cornell.edu/projects/kddcup/datasets.html}};
these have been used to evaluate algorithms in the context of
modularity based clustering~\cite{dnt-aaaac-13,dyt-t-09,ndyt-a-11}.





\paragraph{Small examples.} 
Many publications about static graph clustering include the analysis
of small networks to illustrate some properties of the clusterings
produced by their algorithm.  A famous example for that is the karate
network collected by Zachary in 1977~\cite{z-ifmcf-77}, which
describes friendship links between members of a karate club before the
club split up due to an internal dispute; a typical question is
whether a clustering algorithm is able to predict the split given the
network structure.  Usually these networks are small enough to be
visualized entirely in an article, which enables readers to compare
different clusterings of these networks across several publications.
The closest evolving analog to Zachary's karate network is the
Southern Women data set collected in 1933 by Davis et
al.~\cite{dgg-ds-41}.  It contains data on the social activities of 18
women observed over a nine-month period.  Within this period, they
recorded for 14 informal events whether these women participated or
not.  It has been used as a test set by Berger et
al.~\cite{bkt-affci-07}, Berger-Wolf and Saia~\cite{bs-afads-06}, and
Yang et al.~\cite{yczj-db-11}.  Another interesting small example is
Grevi's zebra data set~\cite{sfd-ng-07} used by Berger et
al.~\cite{bkt-affci-07}.  It consists of information about the spatial
proximity of members of a zebra herd observed over three months,
corresponding to 44 observations or time steps.

\paragraph{Other Data Sets.}
In the following, we will list some further sources for dynamic graph
data already used to evaluate dynamic graph clustering
algorithms. 
Xie et al.~\cite{xcs-lrtic-13} have used graphs representing the
topology of the internet at the level of autonomous systems (AS
Graphs) based on data collected by the University of Oregon Route
Views Project~\cite{lkf-gotdl-05}. These data are available from the
Stanford Large Network Dataset
Collection\footnote{\url{http://snap.stanford.edu/data/index.html}}.
Xu et al.~\cite{xki-tcdsn-11} try to identify communities of spammers
in data from Project Honey
Pot\footnote{\url{http://www.projecthoneypot.org/}}, an ongoing
project to identify spammers.  Sun et al.~\cite{sthgz-c-10} use data
extracted from the social bookmarking web service
Delicious\footnote{\url{https://delicious.com/}}, which naturally
comes with a plenitude of metadata.  Kim et al.~\cite{kmm-m-10} use
data from youtube
crawls\footnote{\url{http://netsg.cs.sfu.ca/youtubedata/}} in their
evaluation.  Pang et al.~\cite{pcw-a-09} cluster a dynamic network of
players of World of Warcraft, where edges are based on the information
whether they take part in the same group.




%

\paragraph{Static Networks with artificial dynamics.}
Apart from real world data with a naturally given temporal evolution,
it is also possible to artificially incorporate some dynamics into
originally static data.  Riedy et al.~\cite{rb-mcmms-13}, for example,
consider static real world networks that become dynamic by generating
random edge deletions and insertions.

\subsection{Dynamic Graph Generators}\label{sec:dynGen}
Depending on the aim of designing a certain clustering algorithm,
there are good reasons to use synthetic data as well as good reasons
to use \emph{not only} synthetic data for the evaluation.  Synthetic
data means graphs that are artificially generated by the help of a
graph generator.  Given a number of vertices, these generators decide
which vertices are connected by an edge based on the probability of
such an edge.  The edge probabilities are derived for example from a
preferential attachment process, where vertices that already have a
high degree are connected with higher probability than others, or from
other rules that are characteristic for the particular generator.  In
the context of evolving graphs, graph generators usually not only have
to decide which vertices are linked but also which vertices or edges
are added or deleted.  Furthermore, if the generator incorporates a
hidden ground truth clustering, this usually evolves randomly as well,
which in turn influences edge probabilities.

One reason to include real world instances, i.e., instances that stem from typical applications, in the
experimental evaluation is that they
frequently exhibit very specific properties and symmetries that are
difficult to analyze and rebuild in synthetic data.  
Hence, to predict the
performance of an algorithm in a certain application, using only synthetic data is unrewarding,
since experiments involving
sample instances stemming from this application are often more accurate.

This raises the question of why to use synthetic data at all.  There
are some good arguments that justify the use of synthetic data, at
least together with real world data:
\begin{itemize}
\item Tunable characteristics, as for example the density of the
  generated graphs, allow to evaluate algorithms in detail depending
  on these characteristics. 
  A scenario where this can be useful is when an algorithm yields good results for some networks but bad results on others.
  A study on a large set of generated graphs might help to identify characteristics of the graph that are difficult to handle for the algorithm, which in turn
  might raise some
  potential for improvements.
\item Synthetic graphs can usually be generated in any possible size,
  even very large networks that might not (yet) exist in practice.
  This is especially useful in the context of scalability studies.
\item Using a graph generator, an unlimited number of different
  networks with similar properties can be generated, preventing
  algorithms to focus only on very few
  benchmark instances.  
  This permits to test algorithms on a representative sample of the graph class one is interested in, ensuring
  some degree of significance.
\item In particular in the context of graph clustering, there is another reason why
  synthetic networks are quite popular.  Since there is no general agreement
  on a single objective function evaluating the goodness of the
  clustering,  
  a common approach to evaluate graph
  clusterings independent of any objective function is the comparison
  to a known ground truth clustering.
  The downside of this is that
  real world graphs with a well-motivated ground truth clustering are
  still rare.  For this reason, synthetic networks incorporating a hidden
  ground truth clustering that has been used in the generation process
  are popular.
\end{itemize}

In the following, we aim to give a short overview of models for
synthetic graphs that might be useful in the context of clustering
evolving networks.  We start with some models especially suited for
complex networks that can for example be derived by observing human
interaction, with a particular focus on models that try to explain
their evolution.  In the second part, we give an overview on synthetic
benchmark instances that incorporate a hidden ground-truth clustering,
together with existing approaches to make these benchmarks
dynamic.

Probably the most fundamental model for synthetic graphs are graphs
where every edge exists with a fixed, constant probability, as first
considered by Gilbert~\cite{g-rg-59} in 1959.  Until then, a lot of
effort has been put into alternative models that better capture the
properties of real world complex networks which typically exhibit
characteristics like small diameter, high clustering coefficient and a
powerlaw degree distribution~\cite{n-tsfcn-03}.  Two classical models
are small world networks~\cite{ws-c-98} that explicitly address the
first two issues and the Barab{\'a}si-Albert model~\cite{ba-esrn-99}
that mostly addresses the degree distribution.  The latter can be seen
as a dynamic model for graph growth according to a preferential
attachment process.  Numerous variations thereof exist, most of which
are targeted in capturing more accurately properties observed in real
world social networks~\cite{knt-s-06,lkf-gotdl-05,v-gp-03}.  Leskovec
et al.~\cite{lbkt-m-08} determine automatically, among a set of
parameterized models, the one fitting a set of four large online
networks best based on the associated likelihood values.  Similarly,
Patro et al.~\cite{pdswfk-tmmad-} propose to use an evolutionary
algorithm to choose among a set of parameterized models of network
growth the one fitting a given set of input characteristics best, in
order to automatically learn the best model for different graph
classes.

Although these models incorporate network growth and already reflect
common properties of observed complex networks as for example online
social networks very well, they do not come with a well motivated
inherent ground truth clustering that can be used to evaluate
clustering algorithms.  An exception to this is the model by Zheleva
et at.~\cite{zsg-c-09} that is especially targeted at modeling the
growth of social networks where vertices can additionally choose to
enter special groups of interest.  Here, the assumption is that both
the network and the group structure evolve simultaneously, influencing
each other.  It might be possible to use the group structure chosen by
the vertices as a ground truth clustering for overlapping clusters,
although the group structure is correlated to the network only to a
certain extent.  In the model of Bagrow~\cite{b-e-08}, starting from a
graph generated according to preferential attachment, edges are
randomly rewired to incorporate a given ground truth clustering.
While this approach combines a ground truth clustering with a
realistic degree distribution, the evolution stemming from the
preferential attachment process is lost.


For static graph clustering, two synthetic benchmark sets have been
used very frequently in the literature; the GN benchmark introduced by
Girvan and Newman~\cite{gn-c-02} and the LFR benchmark introduced by
Lancichinetti et al.~\cite{lfr-b-08}.  The GN benchmark is based on
the \emph{planted partition}
model~\cite{bgw-egca-03,ck-agppp-01,ggw-sdgc-07}, also called \emph{ad
  hoc} model, that takes as input a given ground truth clustering and
two parameters $p_\text{in}$ and $p_\text{out}$ that correspond to the
linking probabilities between vertices within the same or different
clusters.  Typically, the GN benchmark is used to determine how well
an algorithm is able to recover the ground truth clustering, depending
on the gap between $p_\text{in}$ and $p_\text{out}$.
The planted partition model has been generalized to
weighted~\cite{flzwd-a-07} and bipartite~\cite{gsa-m-07} graphs as
well as hierarchical~\cite{z-nb-03} and overlapping~\cite{ssa-d-09}
ground truth clusterings.  Closely related are relaxed caveman
graphs~\cite{am-dncss-11,w-ndswp-99}.  Among the dynamic graph
clustering algorithms described here,
\algo{FacetNet}~\cite{lczst-a-09}, the approaches by Yang et
al.~\cite{yczj-db-11} and Kim and Han~\cite{kh-a-09}, and the
algorithm framework by G\"orke et al.~\cite{gmssw-dgccm-11a} have been
evaluated with the aid of planted partition graphs.  The former two
evaluations use graphs from the GN benchmark and introduce dynamics
based on vertex moves; in each time step, a constant fraction of
vertices leave their cluster and move to a random one.  Kim and Han
additionally consider a dynamic network that also incorporates the
forming and dissolving of clusters and vertex addition and deletion.
In contrast to that, G\"orke et al.~use a custom graph generator based
on the planted partition model that introduces dynamics by splitting
and merging clusters in the ground truth
clustering~\cite{gs-agdcr-09}.  In each time step, one edge or vertex
is added or deleted according to the probabilities prescribed by the
current ground truth clustering.  Hence, the actual graph structure
follows the ground truth clustering with some delay.  They also
provide an efficient implementation of this
generator~\cite{gkssw-aegcd-12-2}.

In the LFR benchmark, cluster sizes as well as vertex degrees are
expected to follow a power law distribution.  Similar to the planted
partition model, vertices share a certain fraction of their links with
other vertices in their cluster and the remaining links with random
vertices in other parts of the graph.  The LFR benchmark has been
generalized to weighted and directed graphs, as well as to overlapping
clusters~\cite{lf-b-09}.  Among the clustering algorithms described in
Section~\ref{sec:IncrClus}, Dinh et al.~\cite{dnt-aaaac-13} have used
a modification of this benchmark to a dynamic setting, whereas Cazabet
et al.~\cite{cah-d-10} only use it in a static scenario.  Green et
al.~\cite{gdc-tecds-10} use dynamic benchmarks based on LFR graphs
that incorporate different cluster events, including membership
switching, cluster growth, shrinkage, birth and death, and the merge
and split of clusters.  After the ground truth clustering has been
adapted, a new random graph is drawn according to the mechanisms of
the LFR benchmark, which results in large differences between adjacent
timesteps.

Aldecoa and Mar{\'\i}n~\cite{am-c-12} finally suggest to interpolate
between two graphs with a significant clustering structure by rewiring
edges at random.  This is proposed as an alternative to benchmarks
like the GN or LFR benchmark in the context of static clustering
algorithms.  Here, the assumption is that clusterings of the
intermediate states of the graph during the rewiring process should
have low distance to both the ground truth clustering of the initial
and the final state.  The rewiring process could be seen as a model
for community evolution.
In the context of tracking clusterings over time, Berger et
al.~\cite{bkt-affci-07} do not consider models for dynamic graphs but
two scenarios for the evolution of clusters that are more
sophisticated than random vertex moves or cluster splits and merges.
It remains to mention that, in principle, all generative models used
to infer clusterings via a Bayesian approach discussed in
Section~\ref{sec:IncrClus} might also be used as benchmark instances,
as they naturally come with a dynamic ground truth clustering.

\subsection{Summary}
Nowadays, a lot of large real world networks have been collected and
made available by projects like the Stanford Large Network Dataset
Collection\footnote{\url{http://snap.stanford.edu/data/}}.  One
problem in the context of evaluating clustering algorithms for
evolving networks is that even if the original data itself has a
temporal aspect, this information is often missing in the thereof
constructed networks readily provided in many benchmark sets.  On the
other hand, the listing in Section~\ref{sec:benchmarks} reveals that
there is no real lack of dynamic data that is publicly available.  A
downside of these data is that converting them to dynamic graphs is
often laborious and leaves many degrees of freedom.  As discussed in
the context of the Enron network, data from the same origin can lead
to quite different dynamic graphs, depending on the design choices
taken.  This makes the comparison of results across different
publications cumbersome.  For static graph clustering, a set of very frequently used networks
mostly taken from the websites of
Newman\footnote{\url{http://www-personal.umich.edu/~mejn/netdata/}}
and
Arenas\footnote{\url{http://deim.urv.cat/~aarenas/data/welcome.htm}} gained some popularity in the
orbit of modularity based methods. It would be nice to have a similar set of common
benchmark graphs that are evolving over time.  A related
issue arises in the context of synthetic benchmarks that incorporate a
ground truth clustering.  Although a lot of publications about the
static case exist, there is still no general agreement on how to make
these data dynamic and what realistic dynamic changes in the ground
truth clustering might look like.

\section{Conclusion}
\label{sec:conclusion}
Clustering evolving networks is at least as difficult as clustering
static networks since it inherits all the difficulties from the static
case and is further faced with additional problems that arise from the
evolution of the considered networks.  The difficulties inherited from
static graph clustering are the many different ideas of what a
good clustering is and what a good clustering algorithm is supposed to
do, as well as the absence of approved benchmark instances to evaluate
and compare the performance of clustering algorithms.  Additional
tasks arise whenever we seek for temporal smoothness or want to detect
and visualize the evolution of clusters over time.  Among the vast
number of algorithms designed for detecting clusters in evolving
graphs, in this survey we only considered graph clustering approaches
in online scenarios with an algorithmic focus on the exploitation of
structural information from previous time steps.  We presented several
state-of-the-art algorithms in different categories and summarized the
main features of these algorithms in Table~\ref{tab:properties_dyn}.
As a first step towards common benchmark sets for the evaluation of
clustering algorithms also in evolving networks,
we explicitly listed data and graph generators that were used by the
authors of the publications presented in this survey.  With this list
we aim at viewing the variety of available data and providing a
collection to other authors in order to help them finding reasonable
test instances for their particular algorithm.  Furthermore, we
discussed tasks like cluster mapping, event detection and
visualization, which make the found cluster information beneficial for
further analysis.  We gave a brief overview on state-of-the-art
approaches solving also these problems and gave some further
references where the reader can find more information regarding these
issues.

\bibliographystyle{plain} \bibliography{references_flat}

\begin{thebibliography}{100}

\bibitem{arb-r-12}
Manoj~K. Agarwal, Krithi Ramamritham, and Manish Bhide.
\newblock {Real time discovery of dense clusters in highly dynamic graphs:
  identifying real world events in highly dynamic environments}.
\newblock In {\em Proceedings of the 38th International Conference on Very
  Large Databases (VLDB 2012)}, pages 980--991, 2012.

\bibitem{axy-tcdlh-11}
Charu~C. Aggarwal, Yan Xie, and Philip~S.\ Yu.
\newblock {Towards community detection in locally heterogeneous networks}.
\newblock In {\em Proceedings of the fifth SIAM International Conference on
  Data Mining}, pages 391--402. SIAM, 2011.

\bibitem{azy-a-10}
Charu~C. Aggarwal, Yuchen Zhao, and Philip Yu.
\newblock {A framework for clustering massive graph streams}.
\newblock {\em Statistical Analysis and Data Mining}, 3(6):399--416, 2010.

\bibitem{am-dncss-11}
Rodrigo Aldecoa and Ignacio Mar{\'\i}n.
\newblock {Deciphering Network Community Structure by Surprise}.
\newblock {\em PLoS ONE}, 6:e24195, September 2011.

\bibitem{am-c-12}
Rodrigo Aldecoa and Ignacio Mar{\'\i}n.
\newblock {Closed benchmarks for network community structure characterization}.
\newblock {\em Physical Review~E}, 85:026109, February 2012.

\bibitem{a-e-85}
David~J. Aldous.
\newblock {Exchangeability and related topics}.
\newblock In P.~L. Hennequin, editor, {\em {\'{E}}cole d'{\'{E}}te de
  Probabilit{\'{e}}s de Saint-Flour XIII-1983}, volume 1117 of {\em Lecture
  Notes in Mathematics}, pages 1--198. Springer, 1985.

\bibitem{awf-bsb-92}
Carolyn~J. Anderson, Stanley Wasserman, and Katherine Faust.
\newblock {Building stochastic blockmodels}.
\newblock {\em Social Networks}, 14:137--161, 1992.

\bibitem{asks-d-12}
Albert Angel, Nikos Sarkas, Nick Koudas, and Divesh Srivastava.
\newblock {Dense subgraph maintenance under streaming edge weight updates for
  real-time story identification}.
\newblock {\em Proceedings of the VLDB Endowment}, 5(6):574--585, February
  2012.

\bibitem{adfg-s-07}
Alex Arenas, Jordi Duch, Alberto Fernandez, and Sergio Gomez.
\newblock {Size reduction of complex networks preserving modularity}.
\newblock {\em New Journal of Physics}, 9(176), 2007.

\bibitem{apu-a-09}
Sitaram Asur, Srinivasan Parthasarathy, and Duygu Ucar.
\newblock {An event-based framework for characterizing the evolutionary
  behavior of interaction graphs}.
\newblock {\em ACM Transactions on Knowledge Discovery from Data},
  3(4):16:1--16:36, December 2009.

\bibitem{afgw-cendd-13}
Thomas Aynaud, Eric Fleury, Jean-Loup Guillaume, and Qinna Wang.
\newblock {Communities in Evolving Networks: Definitions, Detection, and
  Analysis Techniques}.
\newblock In Animesh Mukherjee, Monojit Choudhury, Fernando Peruani, Niloy
  Ganguly, and Bivas Mitra, editors, {\em Dynamics On and Of Complex Networks},
  volume~2 of {\em Modeling and Simulation in Science, Engineering and
  Technology}, pages 159--200. Springer, 2013.

\bibitem{ag-s-10}
Thomas Aynaud and Jean-Loup Guillaume.
\newblock {Static community detection algorithms for evolving networks}.
\newblock In {\em Proceedings of the 8th Intl. Symposium on Modeling and
  Optimization in Mobile, Ad Hoc, and Wireless Networks (WiOpt'10)}, pages
  513--519. IEEE Computer Society, 2010.

\bibitem{bhkl-g-06}
Lars Backstrom, Dan Huttenlocher, Jon~M. Kleinberg, and Xiangyang Lan.
\newblock {Group formation in large social networks: membership, growth, and
  evolution}.
\newblock In {\em Proceedings of the 12th ACM SIGKDD International Conference
  on Knowledge Discovery and Data Mining}, pages 44--54. ACM Press, 2006.

\bibitem{b-e-08}
James Bagrow.
\newblock {Evaluating local community methods in networks}.
\newblock {\em Journal of Statistical Mechanics: Theory and Experiment}, page
  P05001, 2008.
\newblock doi:10.1088/1742-5468/2008/05/P05001.

\bibitem{bbp-fcddc-11}
Shweta Bansal, Sanjukta Bhowmick, and Prashant Paymal.
\newblock {Fast Community Detection for Dynamic Complex Networks}.
\newblock In Luciano~F. Costa, Alexandre Evsukoff, G.~Mangioni, and Ronaldo
  Menezes, editors, {\em Complex Networks}, volume 116 of {\em Communications
  in Computer and Information Science}, pages 196--207. Springer, 2011.

\bibitem{ba-esrn-99}
Albert-L{\'a}szl{\'o} Barab{\'a}si and R{\'e}ka Albert.
\newblock {Emergence of scaling in random networks}.
\newblock {\em Science}, 286:509--512, 1999.

\bibitem{bkt-affci-07}
Tanya Berger-Wolf, David Kempe, and C.~Tantipathananandth.
\newblock {A Framework For Community Identification in Dynamic Social
  Networks}.
\newblock In {\em Proceedings of the 13th ACM SIGKDD International Conference
  on Knowledge Discovery and Data Mining}. ACM Press, 2007.

\bibitem{bs-afads-06}
Tanya Berger-Wolf and Jared Saia.
\newblock {A Framework for Analysis of Dynamic Social Networks}.
\newblock In {\em Proceedings of the 12th ACM SIGKDD International Conference
  on Knowledge Discovery and Data Mining}, pages 523--528. ACM Press, 2006.

\bibitem{bs-gp-11}
Charles-Edmond Bichot and Patrick Siarry, editors.
\newblock {\em {Graph Partitioning}}.
\newblock Wiley, 2011.

\bibitem{by-dnemc-08}
Cemal~Cagatay Bilgin and B{\"u}lent Yener.
\newblock {Dynamic Network Evolution: Models, Clustering, Anomaly Detection}.
\newblock Technical report, Rensselaer University, NY, 2008.

\bibitem{bgll-f-08}
Vincent Blondel, Jean-Loup Guillaume, Renaud Lambiotte, and Etienne Lefebvre.
\newblock {Fast unfolding of communities in large networks}.
\newblock {\em Journal of Statistical Mechanics: Theory and Experiment},
  2008(10), 2008.

\bibitem{bms-mhste-11}
Petko Bogdanov, Misael Mongiovi, and Ambuj~K.\ Singh.
\newblock {Mining Heavy Subgraphs in Time-Evolving Networks}.
\newblock In {\em Proceedings of the 2011 IEEE International Conference on Data
  Mining}, pages 81--90. IEEE Computer Society, 2011.

\bibitem{bkw-pmfds-06}
Karsten~M. Borgwardt, Hans-Peter Kriegel, and Peter Wackersreuther.
\newblock {Pattern Mining in Frequent Dynamic Subgraphs}.
\newblock In {\em Proceedings of the 2006 IEEE International Conference on Data
  Mining}, pages 818--822. IEEE Computer Society, 2006.

\bibitem{bdgghnw-omc-08}
Ulrik Brandes, Daniel Delling, Marco Gaertler, Robert G{\"o}rke, Martin
  H{\"o}fer, Zoran Nikoloski, and Dorothea Wagner.
\newblock {On Modularity Clustering}.
\newblock {\em IEEE Transactions on Knowledge and Data Engineering},
  20(2):172--188, February 2008.

\bibitem{bgw-egca-03}
Ulrik Brandes, Marco Gaertler, and Dorothea Wagner.
\newblock {Experiments on Graph Clustering Algorithms}.
\newblock In {\em Proceedings of the 11th Annual European Symposium on
  Algorithms (ESA'03)}, volume 2832 of {\em Lecture Notes in Computer Science},
  pages 568--579. Springer, 2003.

\bibitem{bgw-egcme-07}
Ulrik Brandes, Marco Gaertler, and Dorothea Wagner.
\newblock {Engineering Graph Clustering: Models and Experimental Evaluation}.
\newblock {\em ACM Journal of Experimental Algorithmics}, 12(1.1):1--26, 2007.

\bibitem{bk-facug-73}
Coen Bron and Joep A. G.~M. Kerbosch.
\newblock {Algorithm 457: Finding all cliques of an undirected graph}.
\newblock {\em Communications of the ACM}, 16(9):575--577, 1973.

\bibitem{cbdbhr-la-07}
Umit Catalyurek, Erik Boman, Karen Devine, Doruk Bozdag, Robert Heaphy, and
  Lee~Ann Riesen.
\newblock {Hypergraph-based Dynamic Load Balancing for Adaptive Scientific
  Computations}.
\newblock In {\em 21th International Parallel and Distributed Processing
  Symposium (IPDPS'07)}, pages 1--11. IEEE Computer Society, 2007.

\bibitem{cah-d-10}
R{\'{e}}my Cazabet, Fr{\'{e}}d{\'{e}}ric Amblard, and Chihab Hanachi.
\newblock {Detection of overlapping communities in dynamical social networks}.
\newblock In {\em Proceedings of the 2010 IEEE Second International Conference
  on Social Computing}, pages 309--314. IEEE, 2010.

\bibitem{c-appfg-04}
Deepayan Chakrabarti.
\newblock {AutoPart: Parameter-Free Graph Partitioning and Outlier Detection}.
\newblock In {\em Proceedings of the 8th European Conference on Principles and
  Practice of Knowledge Discovery in Databases}, pages 112--124. ACM Press,
  2004.

\bibitem{ckt-ec-06}
Deepayan Chakrabarti, Ravi Kumar, and Andrew~S. Tomkins.
\newblock {Evolutionary Clustering}.
\newblock In {\em Proceedings of the 12th ACM SIGKDD International Conference
  on Knowledge Discovery and Data Mining}, pages 554--560. ACM Press, 2006.

\bibitem{cfgrstvz-mcmds-10}
Jiyang Chen, Justin Fagnan, Randy Goebel, Reihaneh Rabbany, Farzad Sangi,
  Mansoureh Takaffoli, Eric Verbeek, and Osmar~R. Za{\"{i}}ane.
\newblock {Meerkat: Community Mining with Dynamic Social Networks}.
\newblock In {\em Proceedings in the 10th IEEE International Conference on Data
  Mining - Workshops}, pages 1377--1380. IEEE Computer Society, December 2010.

\bibitem{czg-dclni-09}
Jiyang Chen, Osmar~R. Za{\"{i}}ane, and Randy Goebel.
\newblock {Detecting Communities in Large Networks by Iterative Local
  Expansion}.
\newblock In {\em Proceedings of the 2009 IEEE International Conference on
  Computational Aspects of Social Networks}, pages 105--112. IEEE Computer
  Society, 2009.

\bibitem{cszht-e-07}
Yun Chi, Xiaodan Song, Dengyong Zhou, Koji Hino, and Belle~L. Tseng.
\newblock {Evolutionary spectral clustering by incorporating temporal
  smoothness}.
\newblock In {\em Proceedings of the 13th ACM SIGKDD International Conference
  on Knowledge Discovery and Data Mining}, pages 153--162. ACM Press, 2007.

\bibitem{c-f-05}
Aaron Clauset.
\newblock {Finding local community structure in networks}.
\newblock {\em Physical Review E}, 72(2):026132, August 2005.

\bibitem{cnm-fcsln-04}
Aaron Clauset, Mark E.~J. Newman, and Cristopher Moore.
\newblock {Finding community structure in very large networks}.
\newblock {\em Physical Review E}, 70(066111), 2004.

\bibitem{ck-agppp-01}
Anne Condon and Richard~M. Karp.
\newblock {Algorithms for Graph Partitioning on the Planted Partition Model}.
\newblock {\em Randoms Structures and Algorithms}, 18(2):116--140, 2001.

\bibitem{c-d-89}
George Cybenko.
\newblock {Dynamic load balancing for distributed memory multiprocessors}.
\newblock {\em Journal of Parallel and Distributed Computing}, 7(2):279--301,
  October 1989.

\bibitem{dgg-ds-41}
A.~Davis, B.B. Gardner, and M.~R. Gardner.
\newblock {\em {Deep South}}.
\newblock University of Chicago Press, 1941.

\bibitem{dggw-ecgc-08}
Daniel Delling, Marco Gaertler, Robert G{\"o}rke, and Dorothea Wagner.
\newblock {Engineering Comparators for Graph Clusterings}.
\newblock In {\em Proceedings of the 4th International Conference on
  Algorithmic Aspects in Information and Management (AAIM'08)}, volume 5034 of
  {\em Lecture Notes in Computer Science}, pages 131--142. Springer, June 2008.

\bibitem{dpv-cprn-05}
Imre Der{\'e}nyi, Gergely Palla, and Tam{\'a}s Vicsek.
\newblock {Clique Percolation in Random Networks}.
\newblock {\em Physical Review Letters}, 94:160202, 2005.

\bibitem{dhzgs-amcag-01}
Chris H.~Q. Ding, Xiaofeng He, Hongyuan Zha, Ming Gu, and Horst~D. Simon.
\newblock {A Min-max Cut Algorithm for Graph Partitioning and Data Clustering}.
\newblock In {\em Proceedings of the 2001 IEEE International Conference on Data
  Mining}, pages 107--114. IEEE Computer Society, 2001.

\bibitem{dnt-aaaac-13}
Thang~N. Dinh, N.~P. Nguyen, and My~T. Thai.
\newblock {An Adaptive Approximation Algorithm for Community Detection in
  Dynamic Scale-free Networks}.
\newblock In {\em Proceedings of the 32th Annual Joint Conference of the IEEE
  Computer and Communications Societies (Infocom)}. IEEE Computer Society
  Press, 2013.
\newblock to appear.

\bibitem{dsttz-agami-10}
Thang~N. Dinh, Incheol Shin, Nhi~K. Thai, My~T. Thai, and Taieb Znati.
\newblock {A General Approach for Modules Identification in Evolving Networks}.
\newblock In Michael~J. Hirsch, Panos~M. Pardalos, and Robert Murphey, editors,
  {\em Dynamics of Information Systems}, volume~40 of {\em Springer
  Optimization and Its Applications}, pages 83--100. Springer, 2010.

\bibitem{dt-cdsfn-13}
Thang~N. Dinh and My~T. Thai.
\newblock {Community Detection in Scale-Free Networks: Approximation Algorithms
  for Maximizing Modularity}.
\newblock {\em IEEE Journal on Selected Areas in Communications},
  31(6):997--1006, 2013.

\bibitem{dyt-t-09}
Thang~N. Dinh, Xuan Ying, and My~T. Thai.
\newblock {Towards social-aware routing in dynamic communication networks}.
\newblock In {\em In Proceedings of the 28th International Performance
  Computing and Communications Conference (IPCCC)}, pages 161--168, 2009.

\bibitem{dhw-dhmctc-11}
Christof Doll, Tanja Hartmann, and Dorothea Wagner.
\newblock {Fully-Dynamic Hierarchical Graph Clustering Using Cut Trees}.
\newblock In Frank Dehne, John Iacono, and J{\"o}rg-R{\"u}diger Sack, editors,
  {\em Algorithms and Data Structures, 12th International Symposium (WADS'11)},
  volume 6844 of {\em Lecture Notes in Computer Science}, pages 338--349.
  Springer, August 2011.

\bibitem{dyll-ik-12}
Dongsheng Duan, Yuhua Li, Ruixuan Li, and Zhengding Lu.
\newblock {Incremental K-clique clustering in dynamic social networks}.
\newblock {\em Artificial Intelligence}, 38(2):129--147, August 2012.

\bibitem{ep-r-06}
Nathan Eagle and Alex Pentland.
\newblock {Reality mining: sensing complex social systems}.
\newblock {\em Journal Personal and Ubiquitous Computing}, 10(4):255--268,
  March 2006.

\bibitem{eksx-adbad-96}
Martin Ester, Hans-Peter Kriegel, J{\"o}rg Sander, and Xiaowei Xu.
\newblock {A Density-Based Algorithm for Discovering Clusters in Large Spatial
  Databases with Noise}.
\newblock In {\em Proceedings of the 2nd ACM SIGKDD international conference on
  Knowledge discovery and Data Mining}, pages 226--231. ACM Press, 1996.

\bibitem{eb-aco-98}
Martin~G.\ Everett and Stephen~P. Borgatti.
\newblock {Analyzing clique overlap}.
\newblock {\em Connections}, 21(1):49--61, 1998.

\bibitem{f-cacad-09}
Tanja Falkowski.
\newblock {\em {Community Analysis in Dynamic Social Networks}}.
\newblock PhD thesis, Otto-von-Guericke-Universit{\"{a}}t Magdeburg, 2009.

\bibitem{fbs-mvess-06}
Tanja Falkowski, J{\"o}rg Bartelheimer, and Myra Spiliopoulou.
\newblock {Mining and Visualizing the Evolution of Subgroups in Social
  Networks}.
\newblock In {\em IEEE/WIC/ACM International Conference on Web Intelligence},
  pages 52--58. IEEE, 2006.

\bibitem{fbs-dengr-07}
Tanja Falkowski, Anja Barth, and Myra Spiliopoulou.
\newblock {DENGRAPH: A Density-based Community Detection Algorithm}.
\newblock In {\em IEEE/WIC/ACM International Conference on Web Intelligence},
  pages 112--115. IEEE, 2007.

\bibitem{flzwd-a-07}
Ying Fan, Menghui Li, Peng Zhang, Jinshan Wu, and Zengru Di.
\newblock {Accuracy and precision of methods for community identification in
  weighted networks}.
\newblock {\em Physica~A}, 377(1):363--372, 2007.

\bibitem{ftt-gcmct-04}
Gary~William Flake, Robert~E. Tarjan, and Kostas Tsioutsiouliklis.
\newblock {Graph Clustering and Minimum Cut Trees}.
\newblock {\em Internet Mathematics}, 1(4):385--408, 2004.

\bibitem{f-c-09}
Santo Fortunato.
\newblock {Community detection in graphs}.
\newblock {\em Physics Reports}, 486(3--5):75--174, 2010.

\bibitem{bf-rlcd-07}
Santo Fortunato and Marc Barth{\'e}lemy.
\newblock {Resolution limit in community detection}.
\newblock {\em Proceedings of the National Academy of Science of the United
  States of America}, 104(1):36--41, 2007.

\bibitem{ggw-sdgc-07}
Marco Gaertler, Robert G{\"o}rke, and Dorothea Wagner.
\newblock {Significance-Driven Graph Clustering}.
\newblock In {\em Proceedings of the 3rd International Conference on
  Algorithmic Aspects in Information and Management (AAIM'07)}, Lecture Notes
  in Computer Science, pages 11--26. Springer, June 2007.

\bibitem{gm-addhc-10}
Joachim Gehweiler and Henning Meyerhenke.
\newblock {A Distributed Diffusive Heuristic for Clustering a Virtual P2P
  Supercomputer}.
\newblock In {\em Proc. 7th High-Performance Grid Computing Workshop (HGCW'10)
  in conjunction with 24th Intl. Parallel and Distributed Processing Symposium
  (IPDPS'10)}, pages 1--8. IEEE Computer Society, 2010.

\bibitem{g-rg-59}
Horst Gilbert.
\newblock {Random Graphs}.
\newblock {\em The Annals of Mathematical Statistics}, 30(4):1141--1144, 1959.

\bibitem{gn-c-02}
Michelle Girvan and Mark E.~J. Newman.
\newblock {Community structure in social and biological networks}.
\newblock {\em Proceedings of the National Academy of Science of the United
  States of America}, 99(12):7821--7826, 2002.

\bibitem{gz-tcfat-04}
Peter~A. Gloor and Yan Zhao.
\newblock {TeCFlow - A Temporal Communication Flow Visualizer for Social
  Network Analysis}.
\newblock In {\em ACM CSCW Workshop on Social Networks}, 2004.

\bibitem{gh-mtnf-61}
Ralph~E. Gomory and T.C. Hu.
\newblock {Multi-terminal network flows}.
\newblock {\em Journal of the Society for Industrial and Applied Mathematics},
  9(4):551--570, December 1961.

\bibitem{g-aawsd-10}
Robert G{\"o}rke.
\newblock {\em {An Algorithmic Walk from Static to Dynamic Graph Clustering}}.
\newblock PhD thesis, Fakult{\"a}t f{\"u}r Informatik, February 2010.

\bibitem{ghw-dmctc-09}
Robert G{\"o}rke, Tanja Hartmann, and Dorothea Wagner.
\newblock {Dynamic Graph Clustering Using Minimum-Cut Trees}.
\newblock In Frank Dehne, Marina Gavrilova, J{\"o}rg-R{\"u}diger Sack, and
  Csaba~D. T{\'o}th, editors, {\em Algorithms and Data Structures, 11th
  International Symposium (WADS'09)}, volume 5664 of {\em Lecture Notes in
  Computer Science}, pages 339--350. Springer, August 2009.

\bibitem{ghw-dcc-11}
Robert G{\"o}rke, Tanja Hartmann, and Dorothea Wagner.
\newblock {Dynamic Graph Clustering Using Minimum-Cut Trees}.
\newblock {\em Journal of Graph Algorithms and Applications}, 16(2):411--446,
  2012.

\bibitem{gkssw-aegcd-12-2}
Robert G{\"o}rke, Roland Kluge, Andrea Schumm, Christian Staudt, and Dorothea
  Wagner.
\newblock {An Efficient Generator for Clustered Dynamic Random Networks}.
\newblock In {\em Proceedings of the 1st Mediterranean Conference on
  Algorithms}, pages 219--233. Springer, 2012.

\bibitem{gmssw-dgccm-11a}
Robert G{\"o}rke, Pascal Maillard, Andrea Schumm, Christian Staudt, and
  Dorothea Wagner.
\newblock {Dynamic Graph Clustering Combining Modularity and Smoothness}.
\newblock {\em ACM Journal of Experimental Algorithmics},
  18(1):1.5:1.1--1.5:1.29, April 2013.

\bibitem{gsw-dcgc-11b}
Robert G{\"o}rke, Andrea Schumm, and Dorothea Wagner.
\newblock {Density-Constrained Graph Clustering}.
\newblock In Frank Dehne, John Iacono, and J{\"o}rg-R{\"u}diger Sack, editors,
  {\em Algorithms and Data Structures, 12th International Symposium (WADS'11)},
  volume 6844 of {\em Lecture Notes in Computer Science}, pages 679--690.
  Springer, August 2011.

\bibitem{gs-agdcr-09}
Robert G{\"o}rke and Christian Staudt.
\newblock {A Generator for Dynamic Clustered Random Graphs}.
\newblock Technical report, iti\_wagner, 2009.
\newblock Informatik, Uni Karlsruhe, TR 2009-7.

\bibitem{gs-igpis-06}
Leo Grady and Eric l.~Schwartz.
\newblock {Isoperimetric Graph Partitioning for Image Segmentation}.
\newblock {\em IEEE Transactions on Pattern Analysis and Machine Intelligence},
  28(3):469--475, 2006.

\bibitem{gdc-tecds-10}
Derek Greene, D{\'{o}}nal Doyle, and P{\'{a}}draig Cunningham.
\newblock {Tracking the Evolution of Communities in Dynamic Social Networks}.
\newblock In {\em Proceedings of the 2010 IEEE/ACM International Conference on
  Advances in Social Networks Analysis and Mining}, pages 176--183. IEEE
  Computer Society, 2010.

\bibitem{gsa-m-07}
Roger Guimer{\`a}, Marta Sales-Pardo, and Lu{\'i}s A.~Nunes Amaral.
\newblock {Module identification in bipartite and directed networks}.
\newblock {\em Physical Review E}, 76:036102, September 2007.

\bibitem{hk-av-13}
Pascal Held and Rudolf Kruse.
\newblock {Analysis and Visualization of dynamic clusterings}.
\newblock In {\em Proceedings of the 46th Hawaii International Conference on
  System Sciences}, pages 1385--1393, 2013.

\bibitem{hkks-tecl-04}
John~E. Hopcroft, Omar Khan, Brian Kulis, and Bart Selman.
\newblock {Tracking Evolving Communities in Large Linked Networks}.
\newblock {\em Proceedings of the National Academy of Science of the United
  States of America}, 101:5244--5253, April 2004.

\bibitem{j-t-12}
Paul Jaccard.
\newblock {The distribution of flora in the alpine zone}.
\newblock {\em New Phytologist}, 11(2):37--50, February 1912.

\bibitem{kvv-ocgbs-04}
Ravi Kannan, Santosh Vempala, and Adrian Vetta.
\newblock {On Clusterings: Good, Bad, Spectral}.
\newblock {\em Journal of the ACM}, 51(3):497--515, May 2004.

\bibitem{kmm-m-10}
Keehyung Kim, Robert~Ian McKay, and Byung-Ro Moon.
\newblock {Multiobjective evolutionary algorithms for dynamic social network
  clustering}.
\newblock In {\em Proceedings of the 12th annual conference on Genetic and
  evolutionary computation}, pages 1179--1186. ACM Press, 2010.

\bibitem{kh-a-09}
Min-Soo Kim and Jiawei Han.
\newblock {A particle-and-density based evolutionary clustering method for
  dynamic networks}.
\newblock In {\em Proceedings of the 35th International Conference on Very
  Large Databases (VLDB 2009)}, pages 622--633, 2009.

\bibitem{knt-s-06}
Ravi Kumar, Jasmine Novak, and Andrew~S. Tomkins.
\newblock {Structure and evolution of online social networks}.
\newblock In {\em Proceedings of the 12th ACM SIGKDD International Conference
  on Knowledge Discovery and Data Mining}, pages 611--617. ACM Press, 2006.

\bibitem{lwy-dcdwg-13}
Jian-Huang Lai, Chang-Dong Wang, and Philip Yu.
\newblock {Dynamic Community Detection in Weighted Graph Streams}.
\newblock In {\em Proceedings of the 2013 SIAM International Conference on Data
  Mining}, pages 151--161. SIAM, 2013.

\bibitem{lf-b-09}
Andrea Lancichinetti and Santo Fortunato.
\newblock {Benchmarks for testing community detection algorithms on directed
  and weighted graphs with overlapping communities}.
\newblock {\em Physical Review E}, 80(1):016118, 2009.

\bibitem{lfk-d-09}
Andrea Lancichinetti, Santo Fortunato, and J{\'a}nos Kert{\'e}sz.
\newblock {Detecting the overlapping and hierarchical community structure of
  complex networks}.
\newblock {\em New Journal of Physics}, 11(033015), 2009.

\bibitem{lfr-b-08}
Andrea Lancichinetti, Santo Fortunato, and Filippo Radicchi.
\newblock {Benchmark graphs for testing community detection algorithms}.
\newblock {\em Physical Review E}, 78(4):046110, October 2008.

\bibitem{lc-b-13}
Conrad Lee and P{\'{a}}draig Cunningham.
\newblock {Benchmarking community detection methods on social media data},
  2013.
\newblock preprint, arXiv:1302.0739 [cs.SI].

\bibitem{ln-csdn-08}
E.~A. Leicht and Mark E.~J. Newman.
\newblock {Community Structure in Directed Networks}.
\newblock {\em Physical Review Letters}, 100(11):118703+, March 2008.

\bibitem{lr-m-99}
Frank~Thomson Leighton and Satish Rao.
\newblock {Multicommodity max-flow min-cut theorems and their use in designing
  approximation algorithms}.
\newblock {\em Journal of the ACM}, 46(6):787--832, 1999.

\bibitem{lbkt-m-08}
Jure Leskovec, Lars Backstrom, Ravi Kumar, and Andrew~S. Tomkins.
\newblock {Microscopic evolution of social networks}.
\newblock In {\em Proceedings of the 14th ACM SIGKDD International Conference
  on Knowledge Discovery and Data Mining}, pages 462--470. ACM Press, 2008.

\bibitem{lkf-gotdl-05}
Jure Leskovec, Jon~M. Kleinberg, and Christos Faloutsos.
\newblock {Graphs Over Time: Densification Laws, Shrinking Diameters and
  Possible Explanations}.
\newblock In {\em Proceedings of the 11th ACM SIGKDD International Conference
  on Knowledge Discovery and Data Mining}, pages 177--187. ACM Press, 2005.

\bibitem{lczst-a-09}
Yu-Ru Lin, Yun Chi, Shenghuo Zhu, Hari Sundaram, and Belle~L. Tseng.
\newblock {Analyzing communities and their evolutions in dynamic social
  networks}.
\newblock {\em ACM Transactions on Knowledge Discovery from Data},
  3(2):8:1--8:31, April 2009.

\bibitem{lwp-elcsl-06}
Feng Luo, James~Z. Wang, and Eric Promislow.
\newblock {Exploring Local Community Structures in Large Networks}.
\newblock In {\em IEEE/WIC/ACM International Conference on Web Intelligence},
  pages 233--239. IEEE, 2006.

\bibitem{m-dlbpn-09}
Henning Meyerhenke.
\newblock {Dynamic Load Balancing for Parallel Numerical Simulations Based on
  Repartitioning with Disturbed Diffusion}.
\newblock In {\em 15th International Conference on Parallel and Distributed
  Systems (ICPADS)}, pages 150--157. IEEE, 2009.

\bibitem{mms-a-09}
Henning Meyerhenke, Burkhard Monien, and Thomas Sauerwald.
\newblock {A new diffusion-based multilevel algorithm for computing graph
  partitions}.
\newblock {\em Journal of Parallel and Distributed Computing}, 69(9):750--761,
  2009.

\bibitem{mms-gpdd-09}
Henning Meyerhenke, Burkhard Monien, and Stefan Schamberger.
\newblock {Graph Partitioning and Disturbed Diffusion}.
\newblock {\em Parallel Computing}, 35(10-11):544--569, October 2009.

\bibitem{m-ewlcs-01}
Boris Mirkin.
\newblock {Eleven Ways to Look at the Chi-Squared Coefficient for Contingency
  Tables}.
\newblock {\em The American Statistician}, 55(2):111--120, May 2001.

\bibitem{mmb-dnv-05}
James Moody, Daniel McFarland, and Skye Bender-deMoll.
\newblock {Dynamic Network Visualization}.
\newblock {\em American Journal of Sociology}, 110(4):1206--1241, 2005.

\bibitem{mm-atbmr-08}
Chris Muelder and Kwan-Liu Ma.
\newblock {A Treemap Based Method for Rapid Layout of Large Graphs}.
\newblock In {\em Proceedings of IEEE Pacific Visualization Symposium}, pages
  231--238, 2008.

\bibitem{mm-rglus-08}
Chris Muelder and Kwan-Liu Ma.
\newblock {Rapid Graph Layout Using Space Filling Curves}.
\newblock {\em IEEE Transactions on Visualization and Computer Graphics},
  14(6):1301--1308, 2008.

\bibitem{n-tsfcn-03}
Mark E.~J. Newman.
\newblock {The Structure and Function of Complex Networks}.
\newblock {\em SIAM Review}, 45(2):167--256, 2003.

\bibitem{n-awn-04}
Mark E.~J. Newman.
\newblock {Analysis of Weighted Networks}.
\newblock {\em Physical Review E}, 70(056131):1--9, 2004.

\bibitem{n-dcsn-04}
Mark E.~J. Newman.
\newblock {Detecting Community Structure in Networks}.
\newblock {\em The European Physical Journal~B}, 38(2):321--330, 2004.

\bibitem{ng-fecsn-04}
Mark E.~J. Newman and Michelle Girvan.
\newblock {Finding and evaluating community structure in networks}.
\newblock {\em Physical Review E}, 69(026113):1--16, 2004.

\bibitem{ndyt-a-11}
N.~P. Nguyen, Thang~N. Dinh, Xuan Ying, and My~T. Thai.
\newblock {Adaptive algorithms for detecting community structure in dynamic
  social networks}.
\newblock In {\em Proceedings of the 30th Annual Joint Conference of the IEEE
  Computer and Communications Societies (Infocom)}, pages 2282--2290. IEEE
  Computer Society Press, 2011.

\bibitem{nmcm-edmdg-09}
Vincenzo Nicosia, G.~Mangioni, V.~Carchiolo, and M.~Malgeri.
\newblock {Extending the Definition of Modularity to Directed Graphs with
  Overlapping Communities}.
\newblock {\em Journal of Statistical Mechanics: Theory and Experiment},
  2009(03):p03024 (23pp), 2009.

\bibitem{nxcgh-iscwa-07}
Huazhong Ning, Wei Xu, Yun Chi, Yihong Gong, and Thomas Huang.
\newblock {Incremental Spectral Clustering With Application to Monitoring of
  Evolving Blog Communities}.
\newblock In {\em Proceedings of the 2007 SIAM International Conference on Data
  Mining}, pages 261--272. SIAM, 2007.

\bibitem{nxcgh-i-10}
Huazhong Ning, Wei Xu, Yun Chi, Yihong Gong, and Thomas Huang.
\newblock {Incremental spectral clustering by efficiently updating the
  eigen-system}.
\newblock {\em Pattern Recognition}, 43:113--127, 2010.

\bibitem{og-a-13}
Michael Ovelg{\"o}nne and Andreas Geyer-Schulz.
\newblock {An ensemble learning strategy for graph clustering}.
\newblock In David~A. Bader, Henning Meyerhenke, Peter Sanders, and Dorothea
  Wagner, editors, {\em Graph Partitioning and Graph Clustering: Tenth DIMACS
  Implementation Challenge}, volume 588 of {\em DIMACS Book}, pages 187--206.
  American Mathematical Society, 2013.

\bibitem{pbv-q-07}
Gergely Palla, Albert-L{\'a}szl{\'o} Barab{\'a}si, and Tam{\'a}s Vicsek.
\newblock {Quantifying social group evolution}.
\newblock {\em Nature}, 446:664--667, April 2007.

\bibitem{pdfv-u-05}
Gergely Palla, Imre Der{\'e}nyi, Ill{\'e}s Farkas, and Tam{\'a}s Vicsek.
\newblock {Uncovering the overlapping community structure of complex networks
  in nature and society}.
\newblock {\em Nature}, 435:814--818, 2005.

\bibitem{pcw-a-09}
Sheng Pang, Changija Chen, and Ting Wei.
\newblock {A realtime community detection algorithm: incremental label
  propagation}.
\newblock In {\em First International Conference on Future Information Networks
  (ICFIN'09)}, pages 313--317. IEEE, 2009.

\bibitem{ps-agacp-98}
YoungJa Park and ManSuk Song.
\newblock {A Genetic Algorithm for Clustering Problems}.
\newblock In {\em Proceedings of the 3rd Annual Conference on Genetic
  Programming}, pages 568--575, 1998.

\bibitem{pdswfk-tmmad-}
Rob Patro, Geet Duggal, Emre Sefer, Hao Wang, Darya Filippova, and Carl
  Kingsford.
\newblock {The Missing Models: A Data-Driven Approach for Learning How Networks
  Grow }.
\newblock In {\em Proceedings of the 18th ACM SIGKDD International Conference
  on Knowledge Discovery and Data Mining}, pages 42--50. ACM Press, 2012.

\bibitem{p-o-00}
Karl Pearson.
\newblock {On the criterion that a given system of deviations from the probable
  in the case of a correlated system of variables is such that it can be
  reasonably supposed to have arisen from random sampling}.
\newblock {\em Philosophical Magazine Series 5}, 50(302):157--175, 1900.

\bibitem{pl-cclnu-06}
Pascal Pons and Matthieu Latapy.
\newblock {Computing Communities in Large Networks Using Random Walks}.
\newblock {\em Journal of Graph Algorithms and Applications}, 10(2):191--218,
  2006.

\bibitem{rak-n-07}
Usha~Nandini Raghavan, R{\'e}ka Albert, and Soundar Kumara.
\newblock {Near linear time algorithm to detect community structures in
  large-scale networks}.
\newblock {\em Physical Review E}, 76(3):036106, September 2007.

\bibitem{r-ocecm-71}
William~M. Rand.
\newblock {Objective Criteria for the Evaluation of Clustering Methods}.
\newblock {\em Journal of the American Statistical Association},
  66(336):846--850, December 1971.

\bibitem{rb-mcmms-13}
Jason Riedy and David~A. Bader.
\newblock {Multithreaded Community Monitoring for Massive Streaming Graph
  Data}.
\newblock In {\em Workshop on Multithreaded Architectures and Applications
  (MTAAP 2013)}, 2013.
\newblock to appear.

\bibitem{rmeb-pcdmg-12}
Jason Riedy, Henning Meyerhenke, David Edinger, and David~A. Bader.
\newblock {Parallel Community Detection for Massive Graphs}.
\newblock In Roman Wyrzykowski, Jack Dongarra, Konrad Karczewski, and Jerzy
  Wa{\'{s}}niewski, editors, {\em Parallel Processing and Applied Mathematics},
  volume 7203 of {\em Lecture Notes in Computer Science}, pages 286--296.
  Springer, 2012.

\bibitem{r-m-78}
Jorma Rissanen.
\newblock {Modeling by shortest data description}.
\newblock {\em Automatica}, 14(5):465--471, September 1978.

\bibitem{rn-m-11}
Randolf Rotta and Andreas Noack.
\newblock {Multilevel local search algorithms for modularity clustering}.
\newblock {\em ACM Journal of Experimental Algorithmics}, 16:2.3:2.1--2.3:2.27,
  July 2011.

\bibitem{r-sa-87}
Peter~J. Rousseeuw.
\newblock {Silhouettes: A graphical aid to the interpretation and validation of
  cluster analysis}.
\newblock {\em Journal of Computational and Applied Mathematics}, 20(0):53--65,
  1987.

\bibitem{sm-dagcu-06}
Barna Saha and Pabitra Mitra.
\newblock {Dynamic Algorithm for Graph Clustering Using Minimum Cut Tree}.
\newblock In {\em Proceedings of the Sixth IEEE International Conference on
  Data Mining - Workshops}, pages 667--671. IEEE Computer Society, December
  2006.

\bibitem{sm-dagcu-07}
Barna Saha and Pabitra Mitra.
\newblock {Dynamic Algorithm for Graph Clustering Using Minimum Cut Tree}.
\newblock In {\em Proceedings of the 2007 SIAM International Conference on Data
  Mining}, pages 581--586. SIAM, 2007.

\bibitem{smm-cvnld-13}
Arnaud Sallaberry, Chris Muelder, and Kwan-Liu Ma.
\newblock {Clustering, Visualizing, and Navigating for Large Dynamic Graphs}.
\newblock In Walter Didimo and Maurizio Patrignani, editors, {\em Proceedings
  of the 20th International Symposium on Graph Drawing (GD'12)}, volume 7704 of
  {\em Lecture Notes in Computer Science}, pages 487--498. Springer, 2013.

\bibitem{ssa-d-09}
E.~N. Sawardecker, Marta Sales-Pardo, and Lu{\'i}s A.~Nunes Amaral.
\newblock {Detection of node group membership in networks with group overlap}.
\newblock {\em The European Physical Journal~B}, 67:277--284, 2009.

\bibitem{s-gc-07}
Satu~Elisa Schaeffer.
\newblock {Graph Clustering}.
\newblock {\em Computer Science Review}, 1(1):27--64, August 2007.

\bibitem{sc-emomg-08}
Philipp Schuetz and Amedeo Caflisch.
\newblock {Efficient Modularity Optimization by Multistep Greedy Algorithm and
  Vertex Mover Refinement}.
\newblock {\em Physical Review~E}, 77(046112), 2008.

\bibitem{scch-d-09}
Huawei Shen, Xueqi Cheng, Kai Cai, and Mao-Bin Hu.
\newblock {Detect overlapping and hierarchical community structure in
  networks}.
\newblock {\em Physica A: Statistical Mechanics and its Applications},
  388(8):1706--1712, 2009.

\bibitem{sm-ncis-00}
Jianbo Shi and Jitendra Malik.
\newblock {Normalized Cuts and Image Segmentation}.
\newblock {\em IEEE Transactions on Pattern Analysis and Machine Intelligence},
  22(8):888--905, 2000.

\bibitem{s-slink-73}
R.~Sibson.
\newblock {SLINK: An optimally efficient algorithm for the single-link cluster
  method}.
\newblock {\em The Computer Journal}, 16(1):30--34, January 1973.

\bibitem{us-onpcs-06}
Ji{\u{r}}{\'{i}} {\u{S}}{\'{i}}ma and Satu~Elisa Schaeffer.
\newblock {On the NP-Completeness of Some Graph Cluster Measures}.
\newblock In {\em Proceedings of the 32rd International Conference on Current
  Trends in Theory and Practice of Computer Science (SOFSEM'06)}, Lecture Notes
  in Computer Science, pages 530--537. Springer, 2006.

\bibitem{sn-epsb-97}
Tom~A.B.\ Snijders and Krzysztof Nowicki.
\newblock {Estimation and Prediction of Stochastic Blockmodels for Graphs with
  Latent Block Structure}.
\newblock {\em Journal of Classification}, 14:75--100, 1997.

\bibitem{snts-monic-06}
Myra Spiliopoulou, Irene Ntoutsi, Yannis Theodoridis, and Rene Schult.
\newblock {MONIC: modeling and monitoring cluster transitions}.
\newblock In {\em Proceedings of the 12th ACM SIGKDD International Conference
  on Knowledge Discovery and Data Mining}, pages 706--711. ACM Press, 2006.

\bibitem{sk-sgpld-12}
Isabelle Stanton and Gabriel Kliot.
\newblock {Streaming Graph Partitioning for Large Distributed Graphs}.
\newblock In {\em Proceedings of the 18th ACM SIGKDD International Conference
  on Knowledge Discovery and Data Mining}, pages 1222--1230. ACM Press, 2012.

\bibitem{sm-ehpcd-13}
Christian Staudt and Henning Meyerhenke.
\newblock {Engineering High-Performance Community Detection Heuristics for
  Massive Graphs}.
\newblock In {\em Proceedings of the 2013 International Conference on Parallel
  Processing}. Conference Publishing Services (CPS), 2013.

\bibitem{sypf-gspfm-07}
Jimeng Sun, Philip~S.\ Yu, Spiros Papadimitriou, and Christos Faloutsos.
\newblock {GraphScope: Parameter-Free Mining of Large Time-Evolving Graphs}.
\newblock In {\em Proceedings of the 13th ACM SIGKDD International Conference
  on Knowledge Discovery and Data Mining}, pages 687--696. ACM Press, 2007.

\bibitem{sthgz-c-10}
Yizhou Sun, Jie Tang, Jiawei Han, Manish Gupta, and Bo~Zhao.
\newblock {Community evolution detection in dynamic heterogeneous information
  networks}.
\newblock In {\em Proceedings of the Eighth Workshop on Mining and Learning
  with Graphs}, pages 137--146. ACM Press, 2010.

\bibitem{syh-rchin-09}
Yizhou Sun, Yintao Yu, and Jiawei Han.
\newblock {Ranking-based Clustering of Heterogeneous Information Networks with
  Star Network Schema}.
\newblock In {\em Proceedings of the 15th ACM SIGKDD International Conference
  on Knowledge Discovery and Data Mining}, pages 797--806. ACM Press, 2009.

\bibitem{sfd-ng-07}
Siva~R. Sundaresan, Ilya~R. Fischhoff, and Jonathan Dushoff.
\newblock {Network metrics reveal differences in social organization between
  two fission-fusion species, Grevy's zebra and onager}.
\newblock {\em Oecologia}, 151(1):140--149, 2007.

\bibitem{tfsz-t-11}
Mansoureh Takaffoli, Justin Fagnan, Farzad Sangi, and Osmar~R. Za{\"{i}}ane.
\newblock {Tracking changes in dynamic information networks}.
\newblock In {\em Proceedings of the 2011 IEEE International Conference on
  Computational Aspects of Social Networks}, pages 94--101. IEEE Computer
  Society, 2011.

\bibitem{trz-ilcid-13}
Mansoureh Takaffoli, Reihaneh Rabbany, and Osmar~R. Za{\"{i}}ane.
\newblock {Incremental Local Community Identification in Dynamic Social
  Networks}.
\newblock In {\em Proceedings of the 2013 IEEE/ACM International Conference on
  Advances in Social Networks Analysis and Mining}. IEEE Computer Society,
  2013.
\newblock to appear.

\bibitem{tpsyf-c-08}
HangHang Tong, Spiros Papadimitriou, Jimeng Sun, Philip~S.\ Yu, and Christos
  Faloutsos.
\newblock {Colibri: fast mining of large static and dynamic graphs}.
\newblock In {\em Proceedings of the 14th ACM SIGKDD International Conference
  on Knowledge Discovery and Data Mining}, pages 686--694. ACM Press, 2008.

\bibitem{v-gp-03}
Alexei V{\'a}zquez.
\newblock {Growing network with local rules: Preferential attachment,
  clustering hierarchy, and degree correlations}.
\newblock {\em Physical Review E}, 67:056104, May 2003.

\bibitem{vmcg-oeuif-09}
Bimal Viswanath, Alan Mislove, Meeyoung Cha, and P.~Krishna Gummadi.
\newblock {On the Evolution of User Interaction in Facebook}.
\newblock In {\em Proceedings of the 2Nd ACM Workshop on Online Social
  Networks}, pages 37--42. ACM Press, 2009.

\bibitem{l-atsc-07}
Ulrike von Luxburg.
\newblock {A Tutorial on Spectral Clustering}.
\newblock {\em Statistics and Computing}, 17(4):395--416, December 2007.

\bibitem{ww-ccao-07}
Silke Wagner and Dorothea Wagner.
\newblock {Comparing Clusterings -- An Overview}.
\newblock Technical Report 2006-04, iti\_wagner, 2007.

\bibitem{ww-sbdg-87}
Yuchung~J.\ Wang and George~Y.\ Wong.
\newblock {Stochastic Blockmodels for Directed Graphs}.
\newblock {\em Journal of the American Statistical Association}, 82:8--19,
  1987.

\bibitem{w-ndswp-99}
Duncan~J.\ Watts.
\newblock {Networks, Dynamics, and the Small-World Phenomenon}.
\newblock {\em American Journal of Sociology}, 105:493--527, 1999.

\bibitem{ws-c-98}
Duncan~J.\ Watts and Steven~H.\ Strogatz.
\newblock {Collective dynamics of 'small-world' networks}.
\newblock {\em Nature}, 393(6684):440--442, June 1998.

\bibitem{xcs-lrtic-13}
Jierui Xie, Mingming Chen, and Boleslaw~K. Szymanski.
\newblock {LabelRankT: Incremental Community Detection in Dynamic Networks via
  Label Propagation}.
\newblock {\em CoRR}, abs/1305.2006, 2013.

\bibitem{xs-lrasl-13}
Jierui Xie and Boleslaw~K. Szymanski.
\newblock {LabelRank: A Stabilized Label Propagation Algorithm for Community
  Detection in Networks}.
\newblock {\em CoRR}, abs/1303.0868, 2013.

\bibitem{xki-tcdsn-11}
Kevin~S. Xu, Mark Kliger, and Alfred O.~Hero III.
\newblock {Tracking Communities in Dynamic Social Networks}.
\newblock In {\em Proceedings of the 4th International Conference on Social
  Computing, Behavioral-Cultural Modeling and Prediction}, pages 219--226,
  2011.

\bibitem{xyfa-scana-07}
Xiaowei Xu, Nurcan Yuruk, Zhidan Feng, and Thomas A.~J. Schweiger.
\newblock {SCAN: A Structural Clustering Algorithm for Networks}.
\newblock In {\em Proceedings of the 13th ACM SIGKDD International Conference
  on Knowledge Discovery and Data Mining}, pages 824--833. ACM Press, 2007.

\bibitem{yczj-db-11}
Tianbao Yang, Yun Chi, Shenghuo Zhu, and Rong Jin.
\newblock {Detecting communities and their evolutions in dynamic social
  networks -- a Bayesian approach}.
\newblock {\em Machine Learning}, 82(2):157--189, February 2011.

\bibitem{yyt-scg-06}
Kai Yu, Shipeng Yu, and Volker Tresp.
\newblock {Soft Clustering on Graphs}.
\newblock In {\em Advances in Neural Information Processing Systems 18},
  page~05. MIT Press, 2006.

\bibitem{ys-m-03}
Stella~X. Yu and Jianbo Shi.
\newblock {Multiclass spectral clustering}.
\newblock In {\em Proceedings of the 9th IEEE International Conference on
  Computer Vision}, pages 313--319, 2003.

\bibitem{z-ifmcf-77}
Wayne~W. Zachary.
\newblock {An Information Flow Model for Conflict and Fission in Small Groups}.
\newblock {\em Journal of Anthropological Research}, 33:452--473, 1977.

\bibitem{zy-ogscs-13}
Yuchen Zhao and Philip~S.\ Yu.
\newblock {On Graph Stream Clustering with Side Information}.
\newblock In {\em Proceedings of the seventh SIAM International Conference on
  Data Mining}, pages 139--150. SIAM, 2013.

\bibitem{zsg-c-09}
Elena Zheleva, Hossam Sharara, and Lise Getoor.
\newblock {Co-evolution of social and affiliation networks}.
\newblock In {\em Proceedings of the 15th ACM SIGKDD International Conference
  on Knowledge Discovery and Data Mining}, pages 1007--1016. ACM Press, 2009.

\bibitem{z-nb-03}
Haijun Zhou.
\newblock {Network landscape from a Brownian particle's perspective}.
\newblock {\em Physical Review E}, 67:041908, 2003.

\end{thebibliography}

\end{document}